\documentclass[11pt,a4paper]{article}
\usepackage{amsmath,amssymb,amsfonts,color,scalefnt}
\usepackage{epsfig,amsthm}
\usepackage{soul}
\usepackage[makeroom]{cancel}
\usepackage{mathrsfs}
\usepackage{epstopdf}
\usepackage[utf8]{inputenc}
\usepackage{hyperref}
\usepackage{slashed}
\usepackage{graphicx}
\usepackage{pict2e}
\usepackage{tikz}
\usepackage{cite}
\usepackage{float}
\usepackage[percent]{overpic}
\usepackage{multirow}
\usepackage{physics}
\usepackage{booktabs}
\usepackage{mwe}
\usepackage{units}
\usepackage [autostyle, english = american]{csquotes}
\usepackage{bbm}
\usepackage[normalem]{ulem} % for sout command
\MakeOuterQuote{"}

\newcommand{\beq}{\begin{eqnarray}}
\newcommand{\eeq}{\end{eqnarray}}
\newcommand{\bea}{\begin{eqnarray}}
\newcommand{\eea}{\end{eqnarray}}
\newcommand{\ba}{\begin{eqnarray}}
\newcommand{\ea}{\end{eqnarray}}
\newcommand{\be}{\begin{equation}}
\newcommand{\ee}{\end{equation}}
\newcommand{\fr}{\frac}
\newcommand{\pa}{\partial}
\newcommand{\non}{\nonumber} 
\newcommand{\crn}{\nonumber \\}
\newcommand{\bpmatrix}{\begin{pmatrix}}
\newcommand{\epmatrix}{\end{pmatrix}}
\newcommand{\doublet}[2]{\begin{pmatrix} #1 \\ #2 \end{pmatrix}}
\renewcommand{\braket}[1]{\left(#1\right)}

\newcommand{\citeres}[1]{Refs.\,\cite{#1}}
\newcommand{\gev}{{\text{GeV}}}
\newcommand{\tev}{{\text{TeV}}}
\newcommand{\evv}{{\text{eV}}}
\newcommand{\eff}{{\text{eff}}}

\newcommand{\abbrev}{\scalefont{.9}}
\renewcommand{\eqref}[1]{Eq.\,\,\ref{#1}}
\newcommand{\tab}[1]{Table~\ref{#1}}

\newcommand{\figref}[1]{Fig.\,\ref{#1}}

\newcommand{\oorder}[1]{{\cal O}(#1)}
\newcommand{\calO}[1]{{\cal O}(#1)}
\newcommand{\calM}{{\cal M}}
\newcommand{\calL}{{\cal L}}
\newcommand{\diag}{{\abbrev \text{diag}}}

\renewcommand{\Re}{{\rm Re}}
\renewcommand{\Im}{{\rm Im}}

\newcommand{\ti}[1]{\ensuremath{\tilde{#1}}}

\newcommand{\eg}{{\it e.g. }}

\newcommand{\mueff}{{\mu_{\text{eff}}}}
\newcommand{\comment}[1]{\ignorespaces}

\newcommand{\s}{\newline \vspace*{-3.5mm}}

\newcommand{\sbeta}{{s_{\beta}}}
\newcommand{\cbeta}{{c_{\beta}}}
\newcommand{\tbeta}{{\tan\beta}}

\newcommand{\abslambda}{|\lambda |}
\newcommand{\abskappa}{|\kappa |}
\newcommand{\mhpm}{{M_{H^\pm}^2}}

\newcommand{\cphiy}{c_{\varphi_y}}
\newcommand{\phiy}{{\varphi_y}}

\newcommand{\vs}{v_{s}}

\newcommand{\ReAkappa}{\Re A_{\kappa}}
\newcommand{\ReAlambda}{\Re A_{\lambda}}

\newcommand{\mathcalM}{\mathcal{M}}
\newcommand{\mathcalR}{\mathcal{R}}

\newcommand{\ImAkappa}{\Im A_{\kappa}}
\newcommand{\ImAlambda}{\Im A_{\lambda}}
\newcommand{\phiom}{{\varphi_\omega}}
\newcommand{\ISS}{\text{ISS}}

\newcommand{\tn}{{\tilde{\nu}}}
\newcommand{\tN}{{\tilde{N}}}
\newcommand{\tX}{{\tilde{X}}}
\newcommand{\Usn}{{\mathcal{U}_{\tilde{\nu}}}}

\newcommand{\DRbar}{\overline{\text{DR}}}
\newcommand{\ZH}{\mathcal{R}^H}

\newcommand{\NMSSMCALC}{{\tt NMSSMCALC}\xspace}
\newcommand{\NMSSMCALCnuSS}{{\tt NMSSMCALC-nuSS}\xspace}

\newcommand{\deltaone}{\delta\!}

\newcommand{\email}[1]{\href{mailto:#1}{\nolinkurl{#1}}}

\begin{document}

\renewcommand*{\thefootnote}{\fnsymbol{footnote}}

\begin{flushright}
    KA-TP-16-2025\\
    FR-PHENO-2025-006
\end{flushright}

\begin{center}
	{\LARGE \bfseries Loop-corrected Trilinear Higgs Self-Couplings in the NMSSM with Inverse Seesaw Mechanism \par}

	\vspace{.7cm}
	Thi Nhung
        Dao\textsuperscript{a,}\footnote{\email{nhung.daothi@phenikaa-uni.edu.vn}}, 
Martin Gabelmann\textsuperscript{b,}\footnote{\email{martin.gabelmann@physik.uni-freiburg.de}}, 
Margarete M\"uhlleitner\textsuperscript{c,}\footnote{\email{margarete.muehlleitner@kit.edu}}

	\vspace{.3cm}
	\textit{
		\textsuperscript{a }Phenikaa Institute for Advanced Study, PHENIKAA University, Hanoi 12116, Vietnam\\[.2em]
%		\textsuperscript{b }Deutsches Elektronen-Synchrotron DESY, Notkestr.~85, 22607 Hamburg, Germany\\[.2em]
		\textsuperscript{b }Albert-Ludwigs-Universit\"at Freiburg, Physikalisches Institut, Hermann-Herder-Str.\ 3, 79104 Freiburg, Germany
              \\[.2em]
\textsuperscript{c }Karlsruhe Institute of Techmology, Institute for
Theoretical Physics, Wolfgang-Gaede-Str.\ 1, 76131 Karlsruhe, Germany
	}              
\end{center}

\renewcommand*{\thefootnote}{\arabic{footnote}}
\setcounter{footnote}{0}

\vspace*{0.3cm}
%
%
%\date{\today}

\begin{abstract}
The higher-order corrections  for the SM-like Higgs boson mass and the trilinear Higgs
self-couplings in the Next-to-Minimal Supersymmetric extension of the
Standard Model (NMSSM) with Inverse Seesaw Mechanism are significant and highly correlated. 
We present here the full one-loop corrections to the trilinear Higgs
self-couplings supplemented by the dominant top-Yukawa and strong
coupling induced two-loop corrections from our previous calculations
in the complex NMSSM. These corrections are performed consistently with  
the corresponding Higgs boson mass corrections. We discuss in detail the new effects from the extended 
neutrino and sneutrino sectors on both the trilinear Higgs self-couplings and the SM-like 
Higgs boson mass. When compared to the case of the NMSSM without
Inverse Seesaw Mechanism, the new effects can be up to 10\% for the
effective SM-like trilinear Higgs self-couplings, and up to 4.5\% for the SM-like 
Higgs boson mass for valid parameter points,
  i.e.~points satisfying the Higgs data, the neutrino data, the
constraints from the charged lepton flavor-violating
decays, and the new physics constraints from the oblique parameters
$S, T, U$. The new corrections are also included in the Higgs-to-Higgs
decays for the heavy Higgs states and implemented in the new version
of the Fortran code {\tt NMSSMCALC-nuSS}.    
\end{abstract}

%\maketitle
%\tableofcontents

\thispagestyle{empty}
\vfill
\newpage

%-----------------------------
\section{Introduction}
\label{sec:intro}
%-----------------------------
The Standard Model (SM) has been shown to describe the world of particle physics successfully at 
the highest precision. Moreover, it has been structurally completed with the discovery of the
 Higgs boson by the Large Hadron Collider (LHC) collaborations ATLAS and CMS in 2012 
 \cite{Aad:2012tfa,Chatrchyan:2012xdj}. The subsequent measurements of the properties
  of the discovered particle are consistent with the predictions of 
  the SM \cite{ATLAS:2022vkf,CMS:2022dwd}. 
Despite its success, the SM cannot explain certain phenomena. 
Among these  are the explanation of the tiny non-zero neutrino masses, 
the existence of Dark Matter (DM), and the observed baryon asymmetry of the Universe. 
The SM needs to be extended in both its particle content and  symmetry 
principles in order to address these questions. In the SM, neutrinos 
are massless since only left-handed neutrino components are included. The Higgs 
sector is minimal, consisting of only one  Higgs doublet. There is 
no interaction between the Higgs boson and neutrinos. The Higgs and the 
neutrino sector, hence, do not have an impact on each other. The situation 
changed with the discovery of the neutrino masses~\cite{Super-Kamiokande:1998kpq}. The need to accommodate
 non-zero neutrino masses requires an enlarged neutrino sector, implying 
 interactions between neutrinos and the Higgs boson. In line with the
  Higgs mechanism for the generation of particle masses, the tiny
   neutrino masses would lead to tiny neutrino Yukawa couplings. This is, 
   however, not the case in the so-called inverse seesaw mechanism, 
   \cite{Mohapatra:1986aw,Mohapatra:1986bd}, which allows for possibly large couplings. 
   In this situation, one can expect a significant interplay between the two sectors and noticeable
    impacts of the neutrino sector on the Higgs observables. \s

Supersymmetric theories \cite{Golfand:1971iw,  Volkov:1973ix, Wess:1974tw,
 Fayet:1974pd,Fayet:1977yc, Fayet:1976cr,  Nilles:1982dy,Nilles:1983ge,
   Frere:1983ag,Derendinger:1983bz,Haber:1984rc,  Sohnius:1985qm,Gunion:1984yn, 
   Gunion:1986nh} belong to the most intensely studied and best-motivated theories 
   beyond the SM that potentially solve the most pressing open questions cited above. 
   They require an extended Higgs sector entailing an enlarged Higgs spectrum
    with an interesting phenomenology. Calculations performed within supersymmetry 
    can be transferred to other benchmark models with extended Higgs sectors 
    that are tested by experiment, like e.g.~the 2-Higgs-Doublet Model (2HDM), 
    with a minimum of effort. In supersymmetry, the mass of the SM-like Higgs
     boson is not a free parameter. It is obtained from the supersymmetric input 
     parameters, and higher-order corrections are crucial in this context to 
     obtain a Higgs mass value that is compatible with the 125~GeV mass measured 
     by the LHC experiments (for recent reviews, see \cite{Slavich:2020zjv,R:2021bml}). 
     Precision calculations of the Higgs mass combined with precision measurements
      consequently allow us to restrict the allowed parameter space of supersymmetry 
      and further pin down a more-fundamental theory describing nature. \s

Investigations of the heavy neutrino impact on the SM-like Higgs boson mass have been performed in 
some supersymmetric extensions. In the Minimal Supersymmetric Standard Model (MSSM) 
\cite{Gunion:1989we,Martin:1997ns,Dawson:1997tz,Djouadi:2005gj} extended by the seesaw mechanism, 
the corrections stemming solely from the heavy neutrinos and their superpartners can be several GeV, 
cf.~\cite{Heinemeyer:2010eg,Draper:2013ava,Chun:2014tfa,Gogoladze:2014vea,Biekotter:2017xmf,Biekotter:2019gtq}. 
The MSSM extended by a complex singlet field, the so-called
next-to-Minimal Supersymmetric Standard Model (NMSSM) \cite{Fayet:1974pd,Dine:1981rt,Barbieri:1982eh,Nilles:1982dy,Frere:1983ag,Derendinger:1983bz,Ellis:1988er,Drees:1988fc,Ellwanger:1993xa,Ellwanger:1995ru,Ellwanger:1996gw,Elliott:1994ht,King:1995vk,Franke:1995tc,Maniatis:2009re,Ellwanger:2009dp},
 solves the $\mu-$problem and provides with its non-minimal Higgs sector and larger neutralino 
 sector a rich phenomenology \cite{0908.4231v1,Balazs:2007pf}. The NMSSM, with an extended neutrino sector,
  can have a significant impact on the SM-like Higgs boson mass, as shown in \cite{Gogoladze:2008wz,Gogoladze:2012jp,Wang:2013jya}. Recently, our group has also contributed to the investigation of the Higgs boson mass in the context 
  of extended Higgs sectors in supersymmetric theories. We have calculated in \cite{Dao:2021vqp} the full
   one-loop corrections in the complex NMSSM extended by six singlet
   leptonic superfields together with the dominant two-loop
   corrections computed at ${\cal O}(\alpha_t^2$
   in the limit of the complex MSSM and the  ${\cal O}(\alpha_t
   \alpha_s)$ corrections in the vanilla complex NMSSM. We found that the impact of these 
   singlet superfields on the $h_u$-like Higgs boson\footnote{The Higgs boson with dominant up-type-like doublet component $h_u$ behaves like the SM Higgs.} can be of up to $5\,\gev$, taking into 
    account the constraints from Higgs data, active light neutrino data, oblique parameters, and lepton 
    flavor violation decays such as $\mu\to e\gamma$ and $\tau\to e\gamma$, $\tau\to \mu\gamma$. We have 
    furthermore found significant effects of the extended (s)neutrino sector on the muon/electron 
    anomalous magnetic moment (AMM), as shown in \cite{Dao:2022rui}, where we also computed the impact 
    on the electric dipole moment. All our calculations have been implemented in the public {\tt   Fortran}
     code  \NMSSMCALCnuSS, which computes the Higgs boson masses and mixings as well as 
     the Higgs boson decay widths and branching ratios, taking into
     account the  state-of-the-art higher-order 
     QCD corrections. The code furthermore calculates the electric dipole moment and the anomalous
      magnetic moment of charged leptons, including the full one-loop and the dominant two-loop corrections. \s

 The ultimate confirmation that the Higgs mechanism is responsible for
 mass generation is given by the reconstruction of the Higgs potential
 itself. This requires the measurement of the Higgs self-couplings
 \cite{Djouadi:1999gv,Djouadi:1999rca,Muhlleitner:2000jj}. At
 the LHC the trilinear Higgs coupling (THC) is directly accessible in
 Higgs pair production through gluon fusion at hadron colliders
 \cite{Djouadi:1999rca,DiMicco:2019ngk} and in double Higgs
 strahlung and $W$ boson fusion into Higgs pairs at
 $e^+ e^-$ colliders \cite{Djouadi:1999gv}. The Higgs boson 
 couplings to the SM particles have been measured to be
 SM-like. The current experimental constraints at 95\% CL on
 the THC modifier $\kappa_\lambda$, given by the ratio of the SM-like
 THC in the new physics 
 model $\lambda_{hhh}$ to the tree-level THC
 $\lambda_{\text{SM}}^{(0)}$ of the SM, 
 $\kappa_\lambda\equiv \lambda_{hhh}/\lambda_{\text{SM}}^{(0)}$,  on
 the other hand, are still rather loose, with $-1.2< \kappa_\lambda < 7.2$ given by ATLAS
 \cite{ATLAS:2024ish} and $-1.24< \kappa_\lambda < 6.9$ given by CMS \cite{CMS:2022dwd}, 
 assuming a SM top-quark Yukawa coupling. There is hence still room for
 beyond-SM physics in the THC \cite{Abouabid:2021yvw}.
In the high-luminosity phase of the LHC, 
 it is expected that the sensitivity range can be reduced to $-0.5<
 \kappa_\lambda < 1.6$ \cite{ATL-PHYS-PUB-2022-053}, at high-energy
 linear colliders a 10-20\% accuracy can be reachable, see e.g.~\citeres{Durig:2016jrs,Abramowicz:2016zbo,Roloff:2019crr}, and at the high-luminosity 100 TeV hadron
 collider the THC can be measured with about 3.5-8\% accuracy \cite{Mangano:2020sao}. This requires precise theoretical
  predictions of the THC in the SM as well as models beyond the SM.
Moreover, the Higgs boson mass and the THC are
 intimately related through the Higgs  potential. For the consistent
 interpretation of the experimental results, the predictions for the
 Higgs boson mass and the trilinear Higgs self-coupling hence have to be
 provided by theory at the same precision. Deviations of the THC from
 the SM value lift the destructive interference between the triangle
 and box diagrams contributing to gluon fusion Higgs pair production
 at the LHC. The impact of potentially large higher-order corrections and
 interference effects on the comparison between the experimental
 results and the theoretical predictions for Higgs boson pair
 production at the LHC can therefore be substantial
 \cite{Arco:2022lai,Heinemeyer:2024hxa,Feuerstake:2024uxs}.  \s

 In the SM, the tree-level THC $\lambda_{\text{SM}}^{(0)}$ is uniquely
 determined by the Higgs boson mass $M_H$ as
 $\lambda_{\text{SM}}^{(0)}= 3 M_H^2/v$, where $v \approx 246$~GeV
 denotes the vacuum expectation value. In the on-shell scheme, the dominant
  one-loop corrections \cite{Hollik:2001px} 
  and the dominant two-loop corrections \cite{Senaha:2018xek,Braathen:2019pxr} arising from the
  top-quark Yukawa coupling are of about $-8.5\%$ and $+1.4\%$,
  respectively.
In the MSSM, the one-loop corrections to the effective
trilinear couplings have been provided many years ago in
\cite{Barger:1991ed,Hollik:2001px,Dobado:2002jz}. The
process-dependent corrections to heavy scalar MSSM Higgs decays into a
lighter Higgs pair have been calculated in
\cite{Williams:2007dc,Williams:2011bu}. The two-loop ${\cal
  O}(\alpha_t \alpha_s)$ SUSY-QCD
corrections to the top/stop-loop induced corrections have been made
available within the effective potential approach in
\cite{Brucherseifer:2013qva}. In the NMSSM, we provided the full
one-loop corrections for the CP-conserving NMSSM \cite{Nhung:2013lpa}. They are
sizeable so that the inclusion of the two-loop corrections is
mandatory to reduce the theoretical uncertainties due to missing
higher-order corrections. Consequently, we subsequently calculated the two-loop
${\cal O}(\alpha_t \alpha_s+\alpha_t^2)$ corrections in the limit of vanishing
external momenta in \cite{Muhlleitner:2015dua,Borschensky:2022pfc}, in the CP-violating
NMSSM. The full one-loop corrections to the Higgs-to-Higgs decays and
other on-shell two-body decays were implemented 
 in \cite{Baglio:2019nlc}. For
corrections to the trilinear Higgs self-couplings in
non-supersymmetric (non-SUSY) Higgs models, see for example
Refs.~\cite{Kanemura:2002vm,Kanemura:2004mg,Kanemura:2015mxa,Kanemura:2017wtm,Basler:2017uxn,Basler:2018cwe,Basler:2019iuu,Basler:2020nrq,Basler:2024aaf} for
one-loop and
Refs.~\cite{Senaha:2018xek,Braathen:2019pxr,Braathen:2019zoh,Bahl:2022jnx,Bahl:2023eau,Bahl:2025wzj}
for two-loop results, and
Refs.~\cite{Krause:2016oke,Bojarski:2015kra,Krause:2017mal,Krause:2018wmo,Denner:2018opp,Krause:2019oar,Krause:2019qwe,Azevedo:2021ylf,Egle:2022wmq,Goodsell:2017pdq}
for the process-dependent Higgs-to-Higgs decays at one-loop level. \s

In this paper, we extend the computation of the higher-order
corrections from the masses to the trilinear Higgs self-couplings in
the framework of the complex NMSSM with the inverse see-saw mechanism
as a benchmark model that, as a SUSY model, allows for the solution of
several open problems of the SM and, with its inverse see-saw
mechanism, allows for an explanation of
massive neutrinos. Since the effect of the extended
(s)neutrino sector on the mass can be significant, as we found in \cite{Dao:2021vqp}, where we calculated the full one-loop corrections to the Higgs boson mass consistently combined with the dominant Yukawa and strong coupling induced two-loop corrections, sizable effects on the THC can be expected as well. In a simple extension of the SM including one Dirac heavy neutrino, the impact of the one heavy neutrino on the SM THC has been investigated in \cite{Baglio:2016ijw}. It can be of order $+20\%$ to $+30\%$ if the off-shell Higgs momentum is fixed to $q_H^*=2500\,\gev$. The conclusion is further confirmed in the SM with the inverse seesaw mechanism, provided that the magnitude of the neutrino Yukawa coupling $|Y_\nu|> 3 $ \cite{Baglio:2016bop}. \s

Furthermore, we  quantify the effect of the (s)neutrinos on the THCs  when including
higher-order corrections and investigate their correlation with the previously computed
higher-order corrections to the Higgs boson masses, taking into account
the current constraints from the LHC Higgs data, from active neutrino
data as well as the LHC constraints on the search for electroweakinos,
charged sleptons and sneutrinos. In detail, we compute the full
one-loop corrections from all sectors of the model to  the THCs
taking into account the full dependence on
external momenta. Furthermore, we factorize the dominant corrections from the (s)top sector and the heavy neutrinos and their  superpartners and then include them to the effective trilinear Higgs self-couplings which are defined at the zero external momentum. Additionally, to improve the accuracy of the theoretical predictions for the THCs and the Higgs-to-Higgs decays, we include also the dominant top-Yukawa ($\alpha_t$) and strong coupling ($\alpha_s$) induced two-loop corrections of the order ${\cal O}(\alpha_t(\alpha_s+\alpha_t))$ from our previous calculations \cite{Muhlleitner:2015dua,Borschensky:2022pfc} in the complex NMSSM. Our results are implemented in the new version of the code \NMSSMCALCnuSS.  \s
  
 The outline of the paper is as follows. In Sec.~\ref{sec:HiggsSec}
we introduce the NMSSM with the inverse seesaw mechanism at tree level   
and set our notation. In Sec.~\ref{sec:effTHCs} we describe our computation of the 
electroweak one-loop corrections to the THCs. In the numerical
results presented in Sec.~\ref{sec:num_anal} we analyze the impact of
the (s)neutrino sector on the THCs in comparison with their impact on
the SM-like Higgs boson mass.  Furthermore, we investigate the
electroweak corrections as a function of the relevant (s)neutrino sector
parameters and  discuss scatter plots obtained from  the parameter scan. In
Sec.~\ref{sec:concl} we summarize our results. 

%%%%%%%%%%%%%%%%%%%%%%%%%%%%%%%%%%%%%%%%%%%%%%%%%%%%%%%%%%%%%
%----------------------------------------
\section{The Model at Tree Level}
\label{sec:HiggsSec}
%----------------------------------------
We consider the complex  NMSSM with inverse seesaw mechanism, 
abbreviated as NMSSM-nuSS. Its main differences compared to the usual complex NMSSM are in the neutrino and sneutrino sectors. We have introduced the model
in detail in our previous studies \cite{Dao:2021vqp,Dao:2022rui}. 
In the following, we only give a short introduction into the model and refer
the reader to \cite{Dao:2021vqp,Dao:2022rui} for a detailed
discussion. \s

The NMSSM-nuSS extends the complex NMSSM by including six gauge-singlet chiral superfields $\hat{N}_i$, $\hat X_i$ ($i=1,2,3$) that carry lepton
number. We further impose a discrete 
$\mathbb{Z}_3$ symmetry with a unit charge of $\omega=
e^{i2\pi/3}$ on the NMSSM-nuSS. With the particular $\mathbb{Z}_3$ charge assignment presented
in \cite{Dao:2021vqp} we find the following NMSSM-nuSS superpotential\footnote{In comparison with \cite{Dao:2021vqp,Dao:2022rui} we have changed the notation of $\mu_X$ to $M_X$.} 
\begin{align}
    \mathcal{W}_{\text{NMSSM-nuSS}} =  W_{\text{MSSM}} - \epsilon_{ab} \lambda \hat{S} \hat{H}_d^a
\hat{H}_u^b +  \frac{1}{3} \kappa \hat{S}^3- y_\nu \epsilon_{ab}  \hat{H}_u^a \hat{L}^b \hat{N}^c +  \lambda_X  \hat S \hat X
\hat X + M_X  \hat X  \hat N^c \,, \label{eq:wnmssm-nuss}
\end{align}
where the complex MSSM superpotential is given by
\begin{align}
	\mathcal{W}_{\text{MSSM}} = - \epsilon_{ab} \braket{ y_u \hat{H}_u^a \hat{Q}^b \hat{U}^c -
y_d \hat{H}_d^a \hat{Q}^b \hat{D}^c - y_e \hat{H}_d^a \hat{L}^b
\hat{E}^c }\;.
    \label{eq:wmssm}
\end{align}
In the above expressions, the quark and lepton superfields are denoted by  $\hat{Q}$, 
$\hat{U}$, 
$\hat{D}$ and $\hat{L}$, $\hat{E}$, respectively, whereas
the Higgs doublet
superfields are $\hat{H}_d$, $\hat{H}_u$ and the singlet superfield is
$\hat{S}$. The totally antisymmetric tensor is given by $\epsilon_{12}= \epsilon^{12}=1$.
Charge conjugated fields are denoted by the superscript $c$. Color and generation indices have been omitted. 
The quark and lepton Yukawa couplings $y_d,$ $y_u, y_e$ are chosen to be real and diagonal.
However, the coupling and mass matrices in the neutrino sector,
$y_\nu,\lambda_X, M_X$, are in general non-diagonal and
complex. The NMSSM-specific coupling parameters $\lambda$ and $\kappa$
are complex. The soft SUSY-breaking NMSSM-nuSS Lagrangian
respecting the gauge symmetries 
and the global $\mathbb{Z}_3$ symmetry reads
\begin{align}
\mathcal L_{\text{NMSSM-nuSS}}^{\text{soft}} =& \mathcal L_{\text{NMSSM}}^{\text{soft}}
+ (\epsilon_{ab} y_\nu  A_{\nu}  H_u^a \ti{L}^b\ti N^*  \crn
& + \lambda_X A_X  S\ti X \ti X
+\mu_X B_{\mu_X}  \ti X \ti N^* + h.c.)   - \ti m^2_X \abs{\ti X}^2 - \ti m^2_N \abs{\ti N}^2   \; ,
\end{align}
with the neutrino trilinear couplings $A_\nu, A_X$
and the bilinear mass $B_{\mu_X}$. The soft SUSY-breaking mass
parameters $\ti m^2_X, \ti m^2_N $ are $3\times 3$ matrices and the
soft SUSY-breaking NMSSM Lagrangian is 
\begin{align}\label{eq:breaking_term}\notag
\mathcal{L}_{\rm soft,NMSSM} = & -m_{H_d}^2 H_d^\dagger H_d - m_{H_u}^2
H_u^\dagger H_u -
m_{\tilde{Q}}^2 \tilde{Q}^\dagger \tilde{Q} - m_{\tilde{L}}^2 \tilde{L}^\dagger \tilde{L}
- m_{\tilde{u}_R}^2 \tilde{u}_R^*
\tilde{u}_R - m_{\tilde{d}_R}^2 \tilde{d}_R^* \tilde{d}_R
\nonumber      \\\nonumber
& - m_{\tilde{e}_R}^2 \tilde{e}_R^* \tilde{e}_R - (\epsilon_{ij} [y_e A_e H_d^i
\tilde{L}^j \tilde{e}_R^* + y_d
A_d H_d^i \tilde{Q}^j \tilde{d}_R^* - y_u A_u H_u^i \tilde{Q}^j
\tilde{u}_R^*] + \mathrm{h.c.})      \\
& -\frac{1}{2}(M_1 \tilde{B}\tilde{B} + M_2
\tilde{W}_j\tilde{W}_j + M_3 \tilde{G}\tilde{G} + \mathrm{h.c.}) \\ \nonumber
& - m_S^2 |S|^2 +
(\epsilon_{ij} \lambda
A_\lambda S H_d^i H_u^j - \frac{1}{3} \kappa
A_\kappa S^3 + \mathrm{h.c.}) \;.
\end{align}
We treat all soft SUSY-breaking gaugino mass parameters, 
$M_k$ ($k=1,2,3$), of the bino, 
wino and gluino fields, $\tilde{B}$, $\tilde{W}_l$ ($l=1,2,3$) and
$\tilde{G}$, respectively, as well as 
the soft SUSY-breaking trilinear couplings,
$A_x$ ($x=\lambda,\kappa,u,d,e$), as complex. In  this model, there are only 
two lepton number violating terms, namely
  $\lambda_X \hat{S} \hat{X} \hat{X}$ and $\lambda_X A_X S
  \tilde{X} \tilde{X}$. \s

  The tree-level Higgs potential  is given by
\beq
V_{H}  &=& (|\lambda S|^2 + m_{H_d}^2)H_d^\dagger H_d+ (|\lambda S|^2
+ m_{H_u}^2)H_u^\dagger H_u +m_S^2 |S|^2 \nonumber \\
&& + \frac{1}{8} (g_2^2+g_1^{2})(H_d^\dagger H_d-H_u^\dagger H_u )^2
+\frac{1}{2} g_2^2|H_d^\dagger H_u|^2 \label{eq:higgspotential} \\ 
&&   + |-\epsilon^{ij} \lambda  H_{d,i}  H_{u,j} + \kappa S^2 |^2+
\big[-\epsilon^{ij}\lambda A_\lambda S   H_{d,i}  H_{u,j}  +\frac{1}{3} \kappa
A_{\kappa} S^3+\mathrm{h.c.} \big] \;,
\nonumber
\eeq
where $g_1$ and $g_2$ denote the $U(1)_Y$ and $SU(2)_L$ gauge
couplings, respectively. Electroweak symmetry breaking (EWSB) occurs at the minimum of
 the Higgs potential where the
 three neutral Higgs boson fields acquire non-vanishing VEVs,
\begin{equation}
    H_d = \doublet{\frac{v_d + h_d +i a_d}{\sqrt 2}}{h_d^-}, \,\, 
    H_u = e^{i\varphi_u}\doublet{h_u^+}{\frac{v_u + h_u +i a_u}{\sqrt 2}},\,\,
    S= \frac{e^{i\varphi_s}}{\sqrt 2} (v_s + h_s + ia_s)\, ,
   \label{eq:vevs}
\end{equation}
with the CP-violating phases $\varphi_{u,s}$. After EWSB, the $W$ and $Z$ bosons get masses,
\be
M_W^2= \fr{1}{4} g_2^2 \braket{v_u^2 + v_d^2}\,, \quad M_Z^2= \fr{1}{4} (g_1^2 + g_2^2) 
\braket{v_u^2 + v_d^2} .
\ee 
The two VEVs $v_u$
and $v_d$ are replaced by  $\tan\beta$ and the SM VEV,
\begin{equation}
   \tan\beta = \fr{v_u}{v_d}, \quad v^2 = v_u^2 +v_d^2\,,
   \label{eq:tan_beta}
\end{equation}
and the singlet VEV is related to the effective $\mu$ parameter  by
\be \mueff = \fr{\lambda v_s e^{i\varphi_s}}{\sqrt 2}.\ee 
The Higgs potential can be cast into the form
\beq
V_H & = & V_H^{\mbox{\scriptsize const}}  +  t_{h_d} h_d + t_{h_u} h_u +
t_{h_s} h_s  +  t_{a_{d}} a_d+  t_{a_{u}} a_u 
+  t_{a_{s}} a_s  \\ \non
&& + 
 \frac{1}{2} \phi^{0,T}  {\mathcal{M}_{\phi\phi}} \, \phi^0 +
 \phi^{c,\dagger} {\mathcal{M}_{h^+h^-}} \, \phi^c 
+V_H^{\phi^3} +V_H^{\phi^4}\;,
\eeq
with $\phi^0 \equiv (h_d, h_u, h_s, a_d, a_u, a_s)^T$ and $\phi^c \equiv
((h_d^-)^*,h_u^+)^T$. The six tadpole coefficients are given by
\begin{subequations}
    \label{eq:tads}
\begin{align}
\frac{t_{h_d}}{v_d } &=m_{H_d}^2+\frac{c_{2\beta}
                            M_Z^2}{2}-\frac{\abslambda \tbeta \vs}{2}
                            \left(\abskappa \cphiy \vs-\sqrt{2}
                            \ImAlambda s_{\phiom-\phiy}+\sqrt{2}
                            \ReAlambda c_{\phiom-\phiy}\right)
\nonumber \\
&\hspace*{0.5cm} +\frac{1}{2} \abslambda^2 \left(\sbeta^2 v^2+\vs^2\right)\,\\
\frac{t_{h_u}}{v_u }
&=m_{H_u}^2-\frac{c_{2\beta}M_Z^2}{2}-\frac{\abslambda  \vs}{2 \tbeta} \left(\abskappa \cphiy \vs-\sqrt{2} \ImAlambda s_{\phiom-\phiy}+\sqrt{2} \ReAlambda c_{\phiom-\phiy}\right) \nonumber\\
&\hspace*{0.5cm} +\frac{1}{2} \abslambda^2 \left(\cbeta^2 v^2+\vs^2\right)\\
\frac{t_{h_s}}{v_S} &= m_S^2+\abskappa^2 \vs^2+\frac{\abslambda^2
                      v^2}{2}+\abslambda \cbeta \sbeta v^2
                      \left(\frac{\ImAlambda
                      s_{\phiom-\phiy}-\ReAlambda c_{\phiom-\phiy}
                      }{\sqrt{2} \vs} -\abskappa \cphiy\right)
                      \nonumber \\
&\hspace*{0.5cm}+\frac{\abskappa \vs (\ReAkappa c_\phiom-\ImAkappa s_\phiom )}{\sqrt{2}}\\
\frac{t_{a_d}}{v_d } &= 
\frac{1}{2} \abslambda \vs \left(-\abskappa \vs s_\phiy+\sqrt{2} \ImAlambda c_{\phiom-\phiy}+\sqrt{2} \ReAlambda s_{\phiom-\phiy}\right)\\
t_{a_u} &= \frac{1}{\tbeta} t_{a_d}  \\
t_{a_s} &= \frac{1}{2} \abslambda \cbeta \sbeta v^2 \left(2 \abskappa
          \vs s_\phiy+\sqrt{2} \ImAlambda c_{\phiom-\phiy}+\sqrt{2}
          \ReAlambda s_{\phiom-\phiy}\right) \nonumber \\
&\hspace*{0.5cm}-\frac{\abskappa \vs^2 (\ImAkappa c_{\phiom}+\ReAkappa
  s_\phiom)}{\sqrt{2}} \;,
\end{align}
\end{subequations}
where the
complex parameters $A_\lambda$ and $A_\kappa$ are decomposed into the corresponding
 imaginary and
 real parts. We introduced the short-hand notation
$c_x\equiv\cos x$ and $s_x\equiv \sin x$. The two combinations of phases that enter the tree-level potential are
\begin{subequations}
\begin{align}
\varphi_y &= \varphi_\kappa - \varphi_\lambda + 2\varphi_s - \varphi_u \label{eq:varphiy} \\
\phiom &= \varphi_\kappa + 3\varphi_s\;. \label{eq:varphiom} 
\end{align}
\end{subequations}
Since we require the five independent
tadpoles to vanish at the minimum of the potential, we can use these minimum conditions to eliminate
 a set of five parameters which we choose as  $\{m_{H_d}^2,m_{H_u}^2,m_S^2,\ImAlambda,\ImAkappa\}$. The $6\times 6$ mass matrix for the neutral Higgs bosons, 
 ${\mathcalM}_{\phi\phi}$, and the $2\times 2$ mass matrix
for the charged Higgs bosons, ${\mathcalM}_{h^+h^-}$, can be found in \cite{Dao:2019qaz}.
The triple  Higgs interactions are collected in $V_H^{\phi^3}$, their explicit expressions
can be found in the appendix A of \cite{Muhlleitner:2015dua}. Constant terms and 
 quartic interactions are summarized in
$V_H^{\mbox{\scriptsize const}}$ and $V_H^{\phi^4}$,
respectively. \s

The neutral Higgs mass matrix  ${\mathcalM}_{\phi\phi}$ is diagonalized by an
orthogonal matrix $\mathcal{R}^H$ such that
\begin{subequations} 
\beq
\diag(m_{h_1}^2,m_{h_2}^2,m_{h_3}^2,m_{h_4}^2,m_{h_5}^2, M_{G^0}^2)&=& 
\mathcal{R}^H {\mathcalM}_{\phi\phi} \mathcal{R}^{H,T}\,,\\
  (h_1,h_2,h_3,h_4,h_5,G^0)^T& = & \mathcal{R}^H\, (h_d, h_u,
h_s, a_d, a_u, a_s)^T. \label{eq:Higgsrotation}
\eeq
\end{subequations}
The tree-level masses are in ascending order, $m_{h_1}\le m_{h_2} \le m_{h_3} \le m_{h_4} \le m_{h_5}$. 
In the 't\,Hooft-Feynman gauge, the neutral Goldstone boson
tree-level mass is equal to the $Z$ boson tree-level mass. The charged Higgs mass matrix ${\mathcalM}_{h^+h^-}$
is diagonalized  with an orthogonal matrix  $\mathcalR^{H^\pm}$,
\begin{subequations} 
\beq   
\diag(\mhpm, M^2_{G^\pm} ) &=&  \mathcalR^{H^\pm} \mathcalM_{h^+h^-}
(\mathcalR^{H^\pm})^T,\\
\mathcalR^{H^\pm} &=& \bpmatrix - \cos\beta & \sin \beta \\
\sin\beta & \cos\beta \epmatrix \;,
\eeq
\end{subequations}
where the charged Goldstone boson tree-level
mass is equal to the $W$ boson tree-level mass in the
't\,Hooft-Feynman gauge. \s

We chose the set of independent parameters entering the Higgs potential  as
\be 
\left\{ t_{h_d},t_{h_u},t_{h_s},t_{a_d},t_{a_s},M_{H^\pm},M_W, M_Z,
e,\tan\beta,|\lambda|,\mu_{\text{eff}},|\kappa|,\ReAkappa,\varphi_\lambda,\varphi_\kappa,\varphi_u,\varphi_s
\right\} \,, \label{eq:inputset1}
\ee
or alternatively with $\ReAlambda$ instead of the charged Higgs mass as input
\be 
\left\{ t_{h_d},t_{h_u},t_{h_s},t_{a_d},t_{a_s},M_W, M_Z,
e,\tan\beta,|\lambda|,\mu_{\text{eff}},|\kappa|,\ReAlambda,\ReAkappa,\varphi_\lambda,\varphi_\kappa,\varphi_u,\varphi_s
\right\}\,. \label{eq:inputset2}
\ee

The gauge, quark and charged lepton sectors of the NMSSM-nuSS are similar to the SM.
Likewise, the neutralino, chargino, up squark, down squark, charged slepton sectors are similar to the ones of the complex NMSSM. We review here the neutrino and sneutrino sectors
which we focus on in this study. With the NMSSM-nuSS superpotential in \eqref{eq:wnmssm-nuss},
one can derive the mass  matrix describing the mixing between the three left-handed neutrino
components, $\nu_L$, with the fermionic fields of the six leptonic singlet superfields, $ \hat{N}_i^c, \hat X_i$, $i=1,2,3$, 
  \be 
\calL_{\text{mass}}^\nu =-\fr{1}{2} \bpmatrix \nu_L & N^c & X
\epmatrix  M_\ISS^\nu  \bpmatrix \nu_L\\ N^c\\ X \epmatrix ,
\ee
where $\nu_L, N^c$ and $X$ are left-handed Weyl spinors and
 \be M_{\ISS}^\nu = \bpmatrix  0 & M_D & 0 \\
M_D^T & 0 & M_X \\
0  & M_X^T & \mu_X \epmatrix . \label{eq:neumassmatrix} \quad \ee 
The mass blocks $M_D, M_X$ and $\mu_X$ are $3\times 3$ matrices and
\be 
M_D =\fr{v_u e^{i\varphi_u}}{\sqrt{2}} y_\nu \,, \quad   \mu_X =
\fr{v_se^{i\varphi_s}}{\sqrt{2}} (\lambda_X + \lambda_X^T) \,. \label{eq:lambdaxinmasses}
\ee
In the inverse seesaw mechanism, one assumes that there is a hierarchy in the eigenvalues
of the matrices  $\mu_X , M_D$ and $M_X$, 
$\abs{m^{\mu_X}}\ll \abs{m^{M_D}} \ll \abs{m^{M_X}}$, so that neutrino masses around $\evv$ can be obtained with $M_X$ elements of $\oorder{\tev}$ and $y_\nu\sim \oorder{1}$.
With this mass hierarchy, the $3\times 3$ light neutrino mass
matrix can be approximated at leading order as 
\be 
M_{\text{light}} = M_D M_N^{-1} M_D^T\,, \; \mbox{ with } \; M_N= M_X \mu_X^{-1}
M_X^T\,,\label{eq:mlight}
\ee
and  is diagonalized by  the Pontecorvo-Maki-Nakagawa-Sakata (PMNS) matrix $U_{\text{PMNS}}$,
 \beq
  U_{\text{PMNS}}^* M_{\text{light}} U_{\text{PMNS}}^\dagger = m_\nu \;, \label{eq:LOexpansion} \quad 
  m_\nu= \diag({m_{\nu_1}},{m_{\nu_2}},{m_{\nu_3}}) \,.
\eeq 
Neutrino oscillation data  constrain the neutrino masses (or relations among them) and the mixing angles entering $U_{\text{PMNS}}$. In order to be able to use these experimental results as input, we re-parameterize the neutrino sector accordingly.
There are two common parameterizations in the literature. Both parametrizations  make use of \eqref{eq:LOexpansion}  such that the light neutrino oscillation data can be taken as input parameters. 
The three complex matrices $M_D, \mu_X, M_X$
are then  not entirely independent, but
one of them will be computed from the others. In the Casas-Ibarra parameterization \cite{Casas:2001sr}, 
$M_D$ is computed from the
 relation
\beq 
M_D =  U_{\text{PMNS}}^T \sqrt{m_\nu}  R  \sqrt{{\calM}_N} V_\nu, \label{eq:MDdefinition} \quad 
{\calM}_N=\diag({M_{N_1}},{M_{N_2}},{M_{N_3}})=  V^*_\nu M_N
V^\dagger_\nu \,,
\eeq
with $R$ being a complex orthogonal matrix parameterized by three complex angles $(\ti\theta_{1,2,3})$, and $V_\nu$ is a unitary matrix diagonalizing $M_N$. One can then obtain $y_\nu$ from \eqref{eq:lambdaxinmasses}. In this parameterization, the set of input parameters
in the neutrino sector reads
%\beq
%{m_{\nu_1}},{m_{\nu_2}},{m_{\nu_3}},\underbrace{\theta_{12},\theta_{23},\theta_{13},\delta_{CP}}_{U_{\text{PMNS}}},
%\underbrace{\theta_{1},\theta_{2},\theta_{3} }_{R}, M_X, \mu_X, A_\nu, m_{\ti L},m_{\ti X},m_{\ti N}, B_{\mu_X}  
%\,. \label{eq:inputcasa}
%\eeq
\beq
{m_{\nu_1}},{m_{\nu_2}},{m_{\nu_3}},{\theta_{12},\theta_{23},\theta_{13},\delta_{CP}},
{\ti\theta_{1},\ti\theta_{2},\ti\theta_{3} }, M_X, \mu_X  
\,. \label{eq:inputcasa}
\eeq
The second possibility used in this work is  the
$\mu_X$-parameterization \cite{Arganda:2014dta} in which $\mu_X$ is computed 
from the relation derived from \eqref{eq:mlight} and \eqref{eq:LOexpansion},
\beq \mu_X= M_X^T M_D^{-1} U_{\text{PMNS}}^* m_\nu  U_{\text{PMNS}}^\dagger  M_D^{T,-1} M_X, 
\eeq
where $M_D$ is calculated from the input $y_\nu$. Therefore the set of input parameters 
in this parameterization is given by 
%\beq
%{m_{\nu_1}},{m_{\nu_2}},{m_{\nu_3}},\underbrace{\theta_{12},\theta_{23},\theta_{13},\delta_{CP}}_{U_{\text{PMNS}}},
%\underbrace{\theta_{1},\theta_{2},\theta_{3} }_{R}, M_X, Y_{\nu}, A_\nu, m_{\ti L},m_{\ti X},m_{\ti N}, B_{\mu_X}  
%\,. \label{eq:inputmux}
%\eeq
\beq
{m_{\nu_1}},{m_{\nu_2}},{m_{\nu_3}},{\theta_{12},\theta_{23},\theta_{13},\delta_{CP}},
M_X, y_{\nu} 
\,. \label{eq:inputmux}
\eeq
After using one of the two parameterizations, we have  fixed all degrees of freedom of the three
complex matrices $M_D, \mu_X, M_X$. We diagonalize the $9\times 9$ neutrino mass matrix in  
\eqref{eq:neumassmatrix} numerically using quadruple precision to obtain a unitary rotation 
matrix $U^\nu$ and nine neutrino mass eigenstates with their masses 
 $m_{n_i}$ $(i=1,...,9)$ being
sorted in ascending order,
\be 
\mathcal{U}_\nu^* M_{\ISS}^\nu \mathcal{U}^\dagger_\nu = \diag(m_{n_1},\cdots,m_{n_9}).  \label{eq:fulldia}
\ee
We define the Majorana neutrino fields as  
\begin{align}
 n_i = \left( \begin{array}{c} \nu_i
\\ \overline{\nu_i} \end{array}
\right) \quad\ \text{with}\quad\ \nu_i = (\mathcal{U}_\nu)_{ik} \nu_{L,k}+ (\mathcal{U}_\nu)_{i(k+3)} N^c_{k}
+ (\mathcal{U}_\nu)_{i(k+6)} X_{k},
\end{align}
where $i=1,...,9$, $k=1,2,3,$ and 
\beq
\overline{\nu_i} = i \sigma_2 \nu_i^{*}.
\eeq
The  Majorana neutrino fields, $n_i$, and their corresponding  mass eigenvalues
 are then used in our actual computation. \s

In the sneutrino sectors, we decompose each complex scalar field into 
its CP-even and CP-odd components,
\begin{subequations}
\begin{align}
    \tn &= \frac{1}{\sqrt{2}} \braket{\tn_+ + i \tn_-}\,,\\
    \tN^* &= \frac{1}{\sqrt{2}} \braket{\tN_+ + i \tN_-}\,,\\
\tX &= \frac{1}{\sqrt{2}} \braket{\tX_+ + i \tX_-}.
\end{align}
\end{subequations}
Due to non-vanishing CP-violating phases, the 9 CP-even components mix with the
9 CP-odd ones resulting in the mass term in the Lagrangian
 \be 
{\cal L}_{\ti\nu}= \fr 12 \psi^T M_{\tn}\psi\;,
\ee
where
$ \psi = (\tn_+,\tN_+,\tX_+,\tn_-,\tN_-,\tX_-)^T $
and $ M_\tn $ is an $18\times 18$ symmetric matrix which can be found 
in \cite{Dao:2021vqp}. We then diagonalize the mass matrix $ M_\tn $
with an orthogonal matrix $
\Usn $
\be
\diag \braket{m^2_{\tilde{n}_1},\cdots,m^2_{\tilde{n}_{18}}} = \Usn M_\tn \Usn^T,
\ee
and order the mass values as $ m^2_{\tilde{n}_1} \leq \cdots \leq
m^2_{\tilde{n}_{18}} $. 
In addition to parameters in the neutrino sector, the sneutrino sector
has the following chosen input parameters 
\be
A_\nu, m_{\ti L},m_{\ti X},m_{\ti N}, B_{\mu_X},
\ee  
which are used to numerically obtain the mass eigenvalues and mixing matrices as described above.
%%%%%%%%%%%%%%%%%%%%%%%%%%%%%%%%%%%%%%%%%%%%%%%%%%%%%%%%%%%%%
\section{Loop Corrections to the Effective THCs and Higgs-to-Higgs Decays \label{sec:effTHCs}}
%%%%%%%%%%%%%%%%%%%%%%%%%%%%%%%%%%%%%%%%%%%%%%%%%%%%%%%%%%%%%
Loop corrections are important for both the Higgs mass predictions
and the triple Higgs couplings. In order to investigate a
 correlation between the two quantities, it is therefore mandatory to 
 perform the calculation of the two using the same 
  framework, \eg renormalization schemes and approximations.
   The loop-corrected THCs considered in this study can be expressed as follows:
 \beq 
 \hat\lambda_{h_i h_j h_k} = \lambda_{h_i h_j h_k}+
 \Delta\lambda_{h_i h_j h_k}^{(1)}+ \Delta\lambda_{h_i h_j h_k}^{(2,\alpha_s\alpha_t)}
 + \Delta\lambda_{h_i h_j h_k}^{(2,\alpha_t^2)}, \quad i,j,k=1,...,5 \;,
 \label{eq:loopTHCs}
 \eeq 
where $ \lambda_{h_i h_j h_k}$ are the tree-level THCs, $\Delta\lambda_{h_i h_j h_k}^{(1)}$
denotes the one-loop corrections  and $\Delta\lambda_{h_i h_j h_k}^{(2,\alpha_s\alpha_t)}$, 
$\Delta\lambda_{h_i h_j h_k}^{(2,\alpha_t^2)}$ are the dominant
two-loop QCD and electroweak corrections, respectively. These two-loop corrections have been 
previously computed in the zero momentum approximation  
in \cite{Muhlleitner:2015dua,Borschensky:2022pfc} for the CP-violating NMSSM without inverse seesaw mechanism.
However, it is possible to embed them into the NMSSM-nuSS in a straight-forward way \cite{Dao:2021vqp}.
Therefore, the approximation in the THCs is the same as the one in the Higgs mass calculation
which includes also the two-loop $\calO{\alpha_s\alpha_t}$ and
$\calO{\alpha_t^2}$ corrections. \s
 
In general, these loop-corrected THCs depend on the external momenta.
 If the decay of the heavier Higgs bosons into 
two lighter Higgs bosons is kinematically allowed, 
we compute the decay width using the loop-corrected THCs
and the on-shell (OS) condition for external Higgs bosons. This means that we set the momentum squared
for each external leg  equal to the loop-corrected Higgs mass squared. To ensure that the OS
condition of the external Higgs bosons is fulfilled, we  take into account the
appropriate wave-function renormalization (WFR) factors \cite{Baglio:2019nlc,Nhung:2013lpa}.
The partial decay width for the decay channel $H_i\to H_jH_k$, where capital $H$ denotes the loop-corrected mass eigenstates and small $h$ the
 tree-level mass eigentates, is then given by
\beq
     \Gamma_{H_i\to H_jH_k}= R_2\fr{\lambda^{1/2}(M_{H_i}^2,M_{H_j}^2,M_{H_k}^2)}
     {16 \pi M_{H_i}^3} \abs{\sum_{i_1,j_1,k_1=1}^5 {\bold Z}^H_{ii_1}{\bold Z}^H_{jj_1}
     {\bold Z}^H_{kk_1} \hat\lambda_{h_i h_j h_k}(M_{H_i}^2,M_{H_j}^2,M_{H_k}^2)}^2. \label{eq:DCwidths}
\eeq
Here $R_2=1/2$ for two identical final states and $R_2=1$ otherwise. $ {\bold Z}$
     denotes the wave-function renormalization factor,  which
      can be found in \cite{Nhung:2013lpa} for
the real case and in \cite{Baglio:2019nlc} for the complex case.
 In the following, we use $M_{H_i}$ ($i=1,...,5$)  to denote the loop-corrected Higgs masses and
$m_{h_i}$  to denote the tree-level ones. All possible decay channels have
been  integrated into the computation of the
total decay widths and decay branching ratios of the heavy Higgs boson
states. \s

The trilinear Higgs self-couplings are not only important for
  Higgs decays, but can also enter e.g.~the
  prediction for Higgs pair production. At the LHC the 
 dominant Higgs pair production process  is given by gluon fusion into Higgs pairs which 
 is a loop induced process. We discuss here only the gluon fusion into two SM-like Higgs bosons, $gg\to H_iH_i$ 
 ($H_i\equiv H_{\text{SM}}$).  At leading order, the Feynman diagrams,
 involving triangle and box contributions, are similar to
the MSSM case \cite{Plehn:1996wb,Dawson:1998py}, except for the ones involving a scalar Higgs boson 
 in the $s$-channel, which gets contributions from five neutral NMSSM
 Higgs bosons in the  
 general CP-violating case. Thus the cross section involves couplings
 of the form $\lambda_{H_{\text{SM}}
   H_{\text{SM}} H_{\text{SM}}}$ and $\lambda_{H_{\text{SM}}
H_{\text{SM}} H_k \ne H_{\text{SM}}}$. Assuming
  that the loop corrections to the gluon fusion process are dominated
  by the corrections to the trilinear Higgs
  self-couplings, which is the case for large corrections to the couplings between the
  Higgs bosons, the loop-corrected trilinear
  Higgs self-couplings can be used in the leading-order gluon fusion
  process in order to obtain an approximate loop-corrected production cross
  section. A proper treatment would require the inclusion of the 
  momentum dependence in the trilinear Higgs self-coupling as well as the
  calculation of notoriously difficult massive double-box diagrams. This  is beyond the purpose of 
  the analysis in
  this paper, where we only want to investigate the effects of the extended (s)neutrino sectors
 on the $gg\to H_{\text{SM}}H_{\text{SM}}$ process.\footnote{For a
   discussion of the momentum effects cf.~e.g.~\cite{Bahl:2023eau,Arco:2025pgx}.}
 Therefore, we simplify our investigation by using the
 effective loop-corrected THCs, which are computed at zero external
 momenta. For this, we include the
 effective loop-corrected THCs that contain the dominant one- and
 two-loop contributions  arising  from the (s)top  sector as in
 \eqref{eq:loopTHCs} and the one-loop contributions 
 from the extended (s)neutrino sector. 
 In the following, we describe in detail our computation of the complete
 one-loop contribution in the NMSSM-nuSS to the triple Higgs couplings 
including the full momentum dependence $\hat\lambda_{h_i h_j
  h_k}^{(1)}(p_1^2,p_2^2,p_3^2)$ and subsequently
  introduce the effective trilinear Higgs self-couplings. 
%%%%%%%%%%%%%%
\subsection{ Complete One-loop Corrections with Full Momentum Dependence }
%%%%%%%%%%%%%% 
The renormalized one-loop corrections to the THCs  can be written in
terms of the unrenormalized one-loop contribution $ \lambda_{h_i h_j h_k}^{(1)}$  and
the counterterm $\delta \lambda_{h_ih_jh_k}^{(1)}$,
\beq
 \Delta\lambda_{h_i h_j h_k}^{(1)} (p^2_1,p_2^2,p_3^2) =  \lambda_{h_i h_j h_k}^{(1)} (p^2_1,p_2^2,p_3^2)+
 \delta \lambda_{h_ih_jh_k}^{(1)},
\eeq
where the $ \lambda_{h_i h_j h_k}^{(1)}$ are computed from genuine
one-loop triangle diagrams generated  with \texttt{FeynArts-3.11} \cite{Kublbeck:1990xc,Hahn:2000kx} and processed with 
\texttt{FormCalc-9.8} \cite{Hahn:formcalc} including full external
momentum dependence. The one-loop counterterms can be written as
\beq 
\delta \lambda_{h_ih_jh_k}^{(1)} &=& \fr{1}2 \sum_{l=1}^6 \bigg(\delta Z_{h_i h_l}\lambda_{h_lh_jh_k}+\delta Z_{h_j h_l}\lambda_{h_ih_lh_k}+
\delta Z_{h_k h_l}\lambda_{h_ih_jh_l}\bigg) \crn
&&+ \sum_{i_1,j_1,k_1=1}^6\ZH_{ii_1}\ZH_{jj_1}\ZH_{kk_1}\delta
\lambda_{i_1j_1k_1}^{hhh} \,,
\eeq    
where we have identified $h_6\equiv G^0$ for simplicity of notation, $\ZH$ is the Higgs rotation matrix defined in \eqref{eq:Higgsrotation} and 
$\delta \lambda_{i_1j_1k_2}^{hhh}$ are the counterterms
in the interaction eigenstates $(h_d,h_u,h_s,a_d,a_u,a_s)$ and are given 
in appendix C of \cite{Muhlleitner:2015dua}. \s

The renormalization scheme applied in this calculation is the mixed
OS-$\DRbar$ scheme as in our Higgs mass calculation
\cite{Dao:2021vqp}. Here, we have the following two options. In the
first option, the input parameters are renormalized as 
\be 
\underbrace{t_{h_d},t_{h_u},t_{h_s},t_{a_d},t_{a_s},M_{H^\pm},M_W, M_Z,
e}_{\text{OS}},\underbrace{\tan\beta,|\lambda|,|\mu_{\text{eff}}|,|\kappa|,\ReAkappa}_{\DRbar}
\,, \label{eq:REinputset1}
\ee 
 where the first nine parameters including five tadpoles, three masses and the 
  electric coupling are renormalized in the OS scheme, and the last five parameters 
  are renormalized in the $\DRbar$ scheme. The complex phases
  $\varphi_\lambda,\varphi_\kappa,\varphi_u,\varphi_s$ do not get
  counterterms at the one-loop order.
  The explicit renormalization conditions for these parameters can be found in 
  \cite{Dao:2021vqp}. 
  In the second option, with $\mbox{Re} (A_\lambda)$
  as input parameter instead of $M_{H^\pm}$, we renormalize the input parameters according to
 \be 
\underbrace{t_{h_d},t_{h_u},t_{h_s},t_{a_d},t_{a_s},M_W, M_Z,
e}_{\text{OS}},\underbrace{\tan\beta,|\lambda|,\mu_{\text{eff}},|\kappa|,\ReAlambda,\ReAkappa}_{\DRbar}
\,. \label{eq:REinputset2}
\ee
We have checked that $\Delta\lambda_{h_i h_j h_k}^{(1)}
(p^2_1,p_2^2,p_3^2) $  are UV finite
and that in the limit of vanishing neutrino Yukawa couplings, our results are in agreement with 
 \cite{Nhung:2013lpa} for the real NMSSM and with
 \cite{Muhlleitner:2015dua} for the complex NMSSM.
 
%%%%%%%%%%%%%%%%%%%%%%%%%%%%%%%%%%%%%%%%%%%%%%%%%%%%%%%%%%%
\subsection{Dominant One-loop Corrections to the Effective THCs}
The effective loop-corrected THCs can be defined as the third derivative of the effective potential with respect to the Higgs fields as
\be
\hat\lambda_{h_i h_j h_k}^{\eff}  = \fr{\pa^2 V_H^{\eff}}{\pa h_i
  \pa h_j \pa h_k} \;,
\ee
which is equivalent to $\hat\lambda_{h_i h_j h_k}$ in Eq.~(\ref{eq:loopTHCs}) evaluated
at zero external momenta.
However, the zero-external-momentum approximation is known to become unreliable
in scenarios that feature very light or even mass-less scalar states. The
singlet-like pseudoscalar state of the NMSSM is a possible candidate for such
a state \cite{Nhung:2013lpa}. Therefore, it is important to organize the calculation to be
numerically robust in scenarios with a light pseudoscalar. While several
workarounds exist in the literature, see e.g. Refs.~\cite{Elias-Miro:2014pca,Martin:2014bca,Kumar:2016ltb,Braathen:2016cqe,Braathen:2017izn}, we simply evade the
problem of infra-red divergences by identifying the dominant contributions to
the effective trilinear coupling and compute
them explicitly at vanishing external momenta. Therefore, all two-point and three-point one-loop integrals will be written in terms of one-loop tadpole integrals.
We compute the effective trilinear couplings assuming that the most dominant contributions are the $\oorder{\alpha_t}$
 corrections originating from the (s)top sector, due to  
 the large top Yukawa coupling, and the (s)neutrino contributions that
 also naturally feature \oorder{1} Yukawa couplings due to the inverse seesaw
mechanism. 
  To further simplify the calculation and to focus on the largest
  corrections, we have performed the calculation in the 
  gaugeless limit where the electric coupling is set to zero ($e\to 0$) but
  the electroweak VEV is kept non-zero
  ($v\ne 0$). For the $\oorder{\alpha_t}$ corrections, the following
  set of parameters has to be renormalized,
  \be t_{h_d}, t_{h_u}, t_{h_s}, t_{a_d}, t_{a_s}, M_{H^\pm}^2, v, \tan\beta, \abs{\lambda},
  \label{eq:alphatOR}
    \ee
    where the first seven parameters are renormalized in the OS scheme
    and the last two parameters are renormalized in the $\DRbar$
    scheme. Explicit expressions of the $\oorder{\alpha_t}$ contributions to 
    the one-loop counterterms, $\delta t_{h_d}, \delta t_{h_u}, \delta t_{h_s}, \delta t_{a_d}, 
    \delta t_{a_s}, \delta M_{H^\pm}^2,$ $ \delta v, \delta\tan\beta,
    \delta\abs{\lambda}$, are given in \cite{Muhlleitner:2015dua}. In
    this study we assume that the neutrino Yukawa couplings are large. 
    Therefore,  the contributions from the (s)neutrino sector can be
     significant. In the following, we present the analytic expressions of the neutrino and 
     sneutrino contributions for the 
     OS counterterms ($\delta t_{h_d}, \delta t_{h_u}, \delta t_{h_s}, \delta t_{a_d}, 
    \delta t_{a_s}, \delta M_{H^\pm}^2,$ $ \delta v$). In the gaugeless limit, 
    the counterterm of the SM
     VEV can be expressed as 
\bea
\fr{\deltaone v}{v} &=&\fr{c_W^2}{2s_W^2}\left(\fr{\deltaone
    M_Z^2}{M_Z^2}-\fr{\deltaone M_W^2}{M_W^2}\right)+\fr12
\fr{\deltaone M_W^2}{M_W^2} \;,
\eea
where 
\bea \fr{\deltaone
  M_W^2}{M_W^2}&=&-\sum_{i=1}^9\fr{m_{n_i}^2\sum_{j=1}^3|\mathcal{U}_{
  \nu_{ij}}|^2}{8\pi^2 v^2}\fr1\epsilon   +\fr{1}{32 \pi^2
  v^2}\bigg[  \sum_{i=1}^{3} \sum_{j=1}^{18} (\mathcal{U}_{\tilde
  n_{ji}}^2 + \mathcal{U}_{\tilde
  \nu_{j(i+9)}}^2) F_0(m_{\tilde n_{j}}^2,m_{\tilde L_i}^2) 
    \\ 
  &&+
        4 \sum_{i=1}^3 A_0(m_{\tilde L_i}^2)   - 2\sum_{i=1}^9 \braket{1- 2 \ln \overline{m_{n_i}^2}
        })m_{n_i}^2 \sum_{j=1}^3 |\mathcal{U}_{\nu_{ij}}|^2   \bigg] 
\\
\fr{\deltaone M_Z^2}{M_Z^2}&=&-\sum_{i=1}^9\fr{m_{n_i}^2\sum_{j=1}^3|\mathcal{U}_{
  \nu_{ij}}|^2}{8\pi^2 v^2}\fr1\epsilon  +\fr{1}{8 \pi^2 v^2}\bigg[\sum_{i=1}^{18}g_{\ti \nu_i \ti \nu_i ZZ} A_0(m_{\ti n_i}^2) +
 4\sum_{i,j=1}^{18} g_{\ti \nu_i\ti \nu_j Z}^2B_{00}( m_{\ti n_i}^2, m_{\ti n_j}^2)
  \bigg] \nonumber\\
  &+&2 \sum_{i,j=1}^9 \bigg[ -2 B_0(m_{n_i}^2,m_{n_j}^2) (
  |g_{\nu_i\nu_jz}^L|^2 + i\leftrightarrow j) 
  - 4 (A_0(m_{n_j}^2) - 2 B_{00}(m_{n_i}^2,m_{n_j}^2) \nonumber \\
  && +
  B_{0}(m_{n_i}^2,m_{n_j}^2)m_{n_i}^2   ) 
  |g_{\nu_i\nu_jz}^L|^2 \bigg]  \eeq 
 where $\ln \overline{m_{n_i}^2}= \ln (m_{n_i}^2/\mu_R^2)$ ($\mu_R$ denotes the
 renormalization scale) and the couplings read
  \bea
g_{\ti \nu_i \ti \nu_j ZZ}  &=& \fr12\sum_{k=1}^3\left( 
 \mathcal{U}_{\ti \nu_{ik} }^*\mathcal{U}_{\ti \nu_{jk} }^*
 + \mathcal{U}_{\ti \nu_{i(k+9)} }^*\mathcal{U}_{\ti \nu_{j(k+9)} }^*  \right)\,, \\ 
 g_{\ti \nu_i \ti \nu_j Z}  &=& \fr12\sum_{k=1}^3\left( 
 \mathcal{U}_{\ti \nu_{i(k+9)} }^*\mathcal{U}_{\ti \nu_{jk} }^*
 - \mathcal{U}_{\ti \nu_{ik} }^*\mathcal{U}_{\ti \nu_{j(k+9)} }^*  \right)\,,\\ 
 g_{ \nu_i  \nu_j Z}^L  &=& \fr12\sum_{k=1}^3 
 \mathcal{U}_{ \nu_{ik} }\mathcal{U}_{ \nu_{jk} }^*  \,.
 \eea
%%%%%%%%%%%%%%%%%%%%%%%%%%%%%%%%%%%%%
The charged Higgs mass counterterm is given by
 \bea
\deltaone M_{H^\pm}^2&=&-\fr{1}{32\pi^2}  g_{H^+\ti n_i\ti n_i H^-} A_0^{\epsilon}(m_{\ti n_i}^2)
- \fr{1}{16\pi^2}g_{H^+\ti l_i\ti l_i H^-} A_0^{\epsilon}(m_{\ti l_i}^2) 
\crn && +
\fr{1}{16\pi^2} \abs{g_{ \ti n_i \ti l_j H^-}}^2B_0^{\epsilon}(m_{\ti l_j}^2, m_{\ti n_i}^2 )
-\fr{1}{8\pi^2} m_{n_i}^2\abs{g^L_{n_i l_j H^+}}^2
B_0^{\epsilon}(0,m_{n_i}^2 ) \,,
\eeq
and the neutral Higgs tadpole counterterms $t_{h_\alpha}$ $(h_\alpha=h_d ,h_u, h_s, a_d,a_s)$ read
\bea
\delta^{\scriptscriptstyle{(1)}}t_{h_\alpha}&=&-\sum_{j=1}^{18}
\fr{g_{h_i \ti n_j\ti n_j}}{32\pi^2}
A_0^{\epsilon}(m_{\ti n_j}^2) 
+ \fr{1}{8\pi^2}\Re[g^{L}_{h_i n_jn_j }] m_{n_j}A_0^{\epsilon}(m_{n_j}^2) \,. 
\eea 
The couplings of the charged Higgs boson to the neutrinos, the charged
leptons, the sneutrinos and the selectrons are given by
%%%%%%
\bea 
g_{H^+\ti n_i\ti n_j H^-}&=&-\frac{1}{2} c_\beta^2 \sum^{3}_{l_1,l_2,l_3} (y_{\nu \
})_{l_1,l_2} (y_{\nu })^*_{l_1,l_3} \bigg[ \mathcal{U}^{\ti \nu  }_{i,l_3+3} \
\left(\mathcal{U}^{\ti \nu }_{j,l_2+3}-i \
\mathcal{U}^{\ti \nu }_{j,l_2+12}\right)+\mathcal{U}^{\ti \nu }_{i,l_3+12} \
\left(\mathcal{U}^{\ti \nu }_{j,l_2+12}+i \
\mathcal{U}^{\ti \nu }_{j,l_2+3}\right)
\crn
&&
+\left(\mathcal{U}^{\ti \nu }_{i,l_2+3}-i \
\mathcal{U}^{\ti \nu }_{i,l_2+12}\right) \left(\mathcal{U}^{\ti \nu }_{j,l_3+3}+i \
\mathcal{U}^{\ti \nu }_{j,l_3+12}\right)\bigg]
+\frac{1}{2} c_\beta \
  s_\beta \sum^{3}_{l_2,l_1} \bigg\{ e^{i \varphi_{u}} \lambda (\lambda \
_X{}^*)_{l_2,l_1} 
\crn
&& \bigg[ -\mathcal{U}^{\ti \nu }_{i,l_2+6} \
\left(\mathcal{U}^{\ti \nu }_{j,l_1+6}-i \
\mathcal{U}^{\ti \nu }_{j,l_1+15}\right) +\mathcal{U}^{\ti \nu }_{i,l_2+15} \
\left(\mathcal{U}^{\ti \nu }_{j,l_1+15}+i \
\mathcal{U}^{\ti \nu }_{j,l_1+6}\right)-\left(\mathcal{U}^{\ti \nu }_{i,l_1+6}-i \
\mathcal{U}^{\ti \nu }_{i,l_1+15}\right) 
\crn
&&
\left(\mathcal{U}^{\ti \nu }_{j,l_2+6}-i \
\mathcal{U}^{\ti \nu }_{j,l_2+15}\right)\bigg]
- e^{-i \varphi_{u}}  \lambda ^c \
(\lambda _X)_{l_2,l_1} \bigg[\mathcal{U}^{\ti \nu }_{i,l_2+6} \
\left(\mathcal{U}^{\ti \nu }_{j,l_1+6}+i \mathcal{U}^{\ti \nu }_{j,l_1+15}\right)
\crn
&&
+i \
\mathcal{U}^{\ti \nu }_{i,l_2+15} \left(\mathcal{U}^{\ti \nu }_{j,l_1+6}+i \
\mathcal{U}^{\ti \nu }_{j,l_1+15}\right)+\left(\mathcal{U}^{\ti \nu }_{i,l_1+6}+i \
\mathcal{U}^{\ti \nu }_{i,l_1+15}\right) \left(\mathcal{U}^{\ti \nu }_{j,l_2+6}+i \
\mathcal{U}^{\ti \nu }_{j,l_2+15}\right)\bigg] \bigg\}\,,
\eea
%%%%%%%%%%
\bea
g_{ \ti n_i \ti l_j H^-}&=&c_\beta \sum^{3}_{l_1,l_2,l_3} \bigg\{\frac{ 1}{\sqrt{2}} e^{-i \varphi_{u}}  \
(\mu_X)_{l_1,l_2} \delta _{j,l_3} \left( \mathcal{U}^{\ti \nu } \
_{i,l_2+6}+i  \mathcal{U}^{\ti \nu } _{i,l_2+15}\right) (y_{\nu \
})^*_{l_3,l_1}
+\frac{1}{2}  v_u (y_{\nu })^*_{l_3,l_1}
 \delta _{j,l_3} (y_{\nu })_{l_2,l_1} 
 \crn
 &&
 \left( \
\mathcal{U}^{\ti \nu *} _{i,l_2}+i  \mathcal{U}^{\ti \nu } _{i,l_2+9}\right) \bigg\}
+\sum^{3}_{l_2,l_1}  \bigg\{ \frac{1}{\sqrt{2}} c_\beta  e^{-i \varphi_{u}} \
 \delta _{j,l_1} (y_\nu A_\nu)_{l_1,l_2} \left( \
\mathcal{U}^{\ti \nu } _{i,l_2+3}+i  \mathcal{U}^{\ti \nu } \
_{i,l_2+12}\right)
\crn
&&
+\frac{1}{2} \lambda  s_\beta v_S (y_{\nu}^*)_{l_1,l_2}  e^{i \varphi_{s}} \delta_{j,l_1} \left( \
\mathcal{U}^{\ti \nu} _{i,l_2+3}+i  \mathcal{U}^{\ti \nu} _{i,l_2+12}\right) \bigg\}\,, 
\eea
%%%%%%%%%%%%%%%%%%%
\bea
g_{H^+\ti l_i\ti l_j H^-} &=&-c_\beta^2 \sum^{3}_{l_1,l_2,l_3} \delta _{j,l_3} \delta_{i,l_2} (y_{\nu })_{l_2,l_1} \
(y_{\nu }^*)_{l_3,l_1} \,,\\
 %%%%%
g^L_{n_i l_j H^+} &=& c_\beta \sum^{3}_{l_2,l_1} e^{i \varphi_{u}} (y_{\nu })_{l_1,l_2} \
\delta_{j,l_1} \mathcal{U}^{\nu}_{i,l_2+3}\,.
\eea
For the contributions from the genuine one-loop triangle
diagrams with neutrinos and
sneutrinos in the loops to the trilinear Higgs-self coupling in the limit
of vanishing momentum we find
\bea  \lambda^{(1, n,\ti n)}_{h_ih_jh_k} &=& -\fr{1}{16\pi^2} C_0(m_{\ti n_{i_1}}^2,m_{\ti n_{i_2}}^2,m_{\ti n_{i_3}}^2) g_{h_i\ti n_{i_1} \ti n_{i_2}}g_{h_j\ti n_{i_1} \ti n_{i_3}}g_{h_k\ti n_{i_2} \ti n_{i_3}} \crn
&+& \fr{1}{32\pi^2}  B_0^{\epsilon}(m_{\ti n_{i_1}}^2,m_{\ti n_{i_2}}^2) \bigg( 
g_{h_i\ti n_{i_1} \ti n_{i_2}} g_{h_j h_k \ti n_{i_1} \ti n_{i_2}}+
g_{h_j\ti n_{i_1} \ti n_{i_2}} g_{h_i h_k \ti n_{i_1} \ti n_{i_2}} +
g_{h_k\ti n_{i_1} \ti n_{i_2}} g_{h_i h_j \ti n_{i_1} \ti n_{i_2}}\bigg) \crn
&+& \fr{1}{4 \pi^2} \bigg[  B_0^{\epsilon}(m_{n_{i_1}}^2,m_{n_{i_2}}^2)\bigg(
\abs{g^{L*}_{i_3i_1h_i} g^{L*}_{i_3i_2h_j} g^{L}_{i_1i_2h_k}  } m_{n_{i_3}}  
  + \abs{g^{L*}_{i_3i_1h_i} g^{L}_{i_3i_2h_k} g^{L}_{i_1i_2h_j}  } m_{n_{i_1}} 
  \crn
  && 
  \abs{g^{L*}_{i_3i_1h_k} g^{L*}_{i_3i_2h_j} g^{L}_{i_1i_2h_i}  } m_{n_{i_3}}   \bigg) 
  + C_0(m_{ n_{i_1}}^2,m_{n_{i_2}}^2,m_{n_{i_3}}^2) m_{n_{i_1}} \bigg(
  \abs{g^{L*}_{i_1i_2h_i} g^{L*}_{i_3i_1h_j} g^{L}_{i_2i_3h_k}  } m_{n_{i_1}}^2 \crn
  && +  \abs{g^{L*}_{i_1i_2h_i} g^{L}_{i_3i_1h_j} g^{L*}_{i_2i_3h_k}  }  m_{n_{i_1}} m_{n_{i_2}}  +  \abs{g^{L}_{i_1i_2h_i} g^{L*}_{i_3i_1h_j} g^{L*}_{i_2i_3h_k}  }  m_{n_{i_1}} m_{n_{i_3}}  \crn
  && +  \abs{g^{L*}_{i_1i_2h_i} g^{L*}_{i_3i_1h_j} g^{L*}_{i_2i_3h_k}  }  m_{n_{i_2}} m_{n_{i_3}}
  \bigg)\bigg]. \eea
  %%%%%%%%%%%%%%  
  The couplings of the neutral Higgs bosons with neutrinos and
  sneutrinos are  derived from the
    following interaction terms in the Lagrangian
  \be  i \bar n_j (g^L_{h_i n_jn_k} P_L + g^{L*}_{h_i n_jn_k} P_R)n_k h_i+ i g_{h_i\ti n_j \ti n_k}h_i\ti n_j \ti n_k  
  + i g_{h_i h_j \ti n_k \ti n_l}h_i h_j \ti n_k \ti n_l\,,\ee
  where 
  \bea g^L_{h_i n_jn_k} &=&-\fr{e^{i \varphi_{s}}}{\sqrt{2}} \left(\mathcal{R}^H_{i,3}+i \mathcal{R}^H_{i,6}\right) 
\sum^{3}_{l_2,l_1=1}(\lambda _X)_{l_1,l_2} \left(
\mathcal{U}^{\nu *}_{3,l_2+6} 
\mathcal{U}^{\nu*}_{k,l_1+6}+ \mathcal{U}^{\nu*}_{3,l_1+6} \
\mathcal{U}^{\nu*}_{k,l_2+6}\right) \crn
&&-\fr{e^{i \varphi_{u}}}{\sqrt{2}} \left(\mathcal{R}^H_{i,2}+i 
\mathcal{R}^H_{i,5}\right) \sum^{3}_{l_2,l_1=1}Y^{\nu }_{l_1,l_2} 
  \left(\mathcal{U}^{\nu *}_{3,l_2+3} \mathcal{U}^{\nu *}_{k,l_1}+\mathcal{U}^{\nu *}_{3,l_1} \mathcal{U}^{\nu *}_{k,l_2+3}\right)\,. \eea
The couplings between the neutral Higgs bosons and sneutrinos are very
lengthy. Therefore we do not display them explicitly
  here, but we will provide them 
upon request. The loop functions appearing in the above expressions
are given by 
  \begin{align}
F_0(x,y) &= -x -y\left( 3-2 \ln\overline{x} - \fr{2x\ln\fr{y}{x}  }{y-x}  \right)\,, \quad & A_0(x) &= x(1 -  \ln\overline{x})\,,\crn
B_{0}(x,y) &= 1-  \ln\overline{y}  - \fr{x\ln \fr{x}{y}}{x-y} \,, \quad \quad 
B_{0}(x,x) = - \ln\overline{x} \,,  \quad &B_{0}(0,x) &=  1-\ln\overline{x},\crn
B_{00}(x,y) &= \fr{x+y}{4} \braket{ \fr32- \ln \overline{y} }  -\fr{x}{4} \fr{\ln\fr{x}{y}}{x-y} \,, \quad \quad \quad 
&B_{00}(x,x) &= \fr{x}{2}\braket{1- \ln\overline{x}} \,,  \crn
B_{00}(0,x) &= \fr{x}{4}\braket{ \fr32
-\ln\overline{x}},\quad \quad A_0^{\epsilon}(x) = \fr{x}{\epsilon} + A_0(x)\,, \quad 
&B_{0}^{\epsilon}(x,y)&= \fr{1}{\epsilon}+ B_{0}(x,y)\,,\\
C_0(x,y,z) &=  \fr{y \ln\fr{x}{y}}{(-x + y) (y - z)} + \fr{
z \ln\fr{x}{z}}{(-x + z) (-y + z)}\,, & C_0(x,x,x) &=-\frac{1}{2 x}
                                                     \,, \crn
 C_0(x,x,z)&=
 \frac{z \ln \left(\frac{x}{z}\right)}{\left(x-z\right)^2}-\frac{1}{x-z}\,,\nonumber
 \end{align}   
 where  $\ln\overline{x} = \ln \fr{x}{\mu_R^2}$ with $\mu_R$ being the renormalization scale.
%%%%%%%%%%%%%%%%%%%%%%%%%%%%%%%%%%%%%%%%%%%%%%%%%%%%%%%%%%%%%
\section{Numerical Analysis \label{sec:num_anal}} 
%%%%%%%%%%%%%%%%%%%%%%%%%%%%%%%%%%%%%%%%%%%%%%%%%%%%%%%%%%%%%
In this section, we analyse the numerical impact of the loop corrections from the (s)neutrino
sector to the Higgs trilinear couplings and their correlation with the
corrections to the SM-like Higgs boson mass. For this purpose, we have performed a parameter scan
where we have taken into account several  constraints from the Higgs data, the neutrino
oscillation data, the oblique parameters $S,T,U$, and the charged
lepton flavor-violating decays. For our parameter scan we proceed as
follows. A given parameter point is first processed by the program
\NMSSMCALCnuSS. The code calculates the loop-corrected Higgs boson
masses in the OS renormalization scheme in the top/stop sector as well
as the tree-level SUSY particle masses together with their mixing
angles as described in Sec.~\ref{sec:HiggsSec}. 
We require one of the neutral CP-even Higgs bosons to be the SM-like
Higgs boson, having a mass in the range  
 \beq \unit[122]{GeV} \le M_{H_{\text{SM}}} \le \unit[128]{GeV} \;. \eeq 
Note that we denote $H_{\text{SM}}$  as
the SM-like Higgs boson, where $H_{\text{SM}}$ can be the lightest scalar
state ${H_1}$ or the next-to-lightest state $H_2$. 
 The masses of the SUSY particles must be larger than the lower bounds 
 from the LEP and LHC searches \cite{ParticleDataGroup:2022pth}, in particular we require
\begin{align}  M_{\chi_1^\pm} & > 94\, \gev, &  M_{\chi^0_2}& > 62.4\, \gev, & M_{\chi^0_3}& > 99.9\, \gev,    \crn
M_{\chi^0_4}& > 116\, \gev, & M_{\ti e_1}& > 107\, \gev, & M_{\ti \mu_1}& > 94\, \gev, \crn
M_{\ti \tau_1}& > 81.9 \, \gev, & M_{\ti \nu_{e/\mu/\tau}}& > 94 \, \gev, & M_{\ti b_1}& > 1270 \, \gev, \crn
M_{\ti t_1}& > 1310 \, \gev, & M_{\ti g}& > 2300\, \gev.  \end{align}
\NMSSMCALCnuSS provides the Higgs decay widths and branching ratios,
including the state-of-the-art higher-order QCD corrections
as well as the effective
Higgs couplings discussed in Sec.~\ref{sec:effTHCs}.
Furthermore, \NMSSMCALCnuSS presents all predictions in the SLHA output format which is passed to \texttt{HiggsTools} \cite{Bahl:2022igd}, containing {\tt HiggsBounds} \cite{Bechtle:2020pkv}, to check if the parameter points pass all the exclusion limits from the searches at LEP, Tevatron, and the LHC, and {\tt HiggsSignals} \cite{Bechtle:2020uwn}, to check if the points are consistent with the LHC data for a 125 GeV Higgs boson within 2$\sigma$ using a $\chi^2$-test.
\NMSSMCALCnuSS~ automatically compares its predictions against
  several experimental results. It can check the neutrino oscillation
data, the oblique parameters $S,T,U$, and the charged lepton
flavor-violating decays and decide whether a given
  parameter point violates any of these constraints. For further information on the use
these constraints, we refer the reader to \cite{Dao:2021vqp} whose
  strategy is followed here as well.  \s

To efficiently identify phenomenologically viable parameter areas, we make use of the Markov Chain Monte Carlo sampling with {\tt EasyScan\_HEP-1.0} \cite{Shang:2023gfy}. In Table~\ref{tab:nmssmscan}, we present the ranges applied for specific parameters appearing in the scan.
\begin{table}
\begin{center}
\begin{tabular}{l|ccc|ccccccc} \toprule
& $t_\beta$ & $\lambda$ & $\kappa$ & $M_1,M_2$  & $A_t$ &
 $m_{\tilde{Q}_3},m_{\tilde{t}_R}$ & $m_{\tilde{\tau}_{R}},m_{\tilde{L}_{3}}$ & $M_{H^\pm}$
& $A_\kappa$ & $\abs{\mueff}$ \\
& & & &  \multicolumn{7}{c}{in TeV} \\ \midrule
min &  1 & 0 & -1 & 0.3  & -5 &  1  & 0.5 & 0.6 &-3 & 0.2 \\
max & 20 & 1 & 1 & 1    & 5  & 3    & 2   & 3   & 3 & 2 \\ \bottomrule
\end{tabular}
\begin{tabular}{l|c|ccccccc} \toprule
& $(y_\nu)_{ij}$ &  $(M_X)_{ii}$  & $(A_\nu)_{ii}$ &
 $(A_x)_{ii}$ & $m_{\tilde{n}_{i}},m_{\tilde{x}_{i}}$ & $(B_{\mu})_i/(A_x)_{ii}$ \\
& & \multicolumn{5}{c}{in TeV} \\ \midrule
min &  -1 &  0.5  & -2 &  -2 &0 & 0.001  \\
max & 1 &   10    & 2  & 2  & 2  & 1    \\ \bottomrule
\end{tabular}
\caption{ The input parameter ranges for the scan. The indices $i,j$ run from 1 to 3. \label{tab:nmssmscan}}
\end{center}
\vspace*{-0.6cm}
\end{table}
Note that we have used the $\mu_X$-parameterization for the neutrino sector.
 All parameters in Table~\ref{tab:nmssmscan}  are chosen to be real in 
the parameter scan and all the $3\times 3$ matrices are set to be diagonal except for the neutrino Yukawa coupling, 
$y_\nu$. We assume that the input parameters $A_{\nu},A_{X},B_{\mu_X}$ 
are identical for all three generations.
The remaining soft-SUSY-breaking parameters are fixed as follows,
\bea
M_3= 2300\,\gev, \quad
m_{\tilde{Q}_{1/2}}=m_{\tilde{L}_{1/2}}=m_{\tilde{x}_R}= 3\,\tev, \quad
A_{b,\tau}= 2\,\tev\,,
\eea
where $x=u,d,c,s,b,e,\mu$. The SM parameters are taken from \cite{ParticleDataGroup:2022pth},
\begin{equation}
\begin{tabular}{lcllcl}
\quad $\alpha$ &=& 1/127.955\,, &\quad $\alpha^{\overline{\mbox{MS}}}_s(M_Z)$ &=&
0.1181\,, \\
\quad $M_W$ &=&80.379~\gev &\quad   $M_Z$ &=& 91.1876~GeV \,, \\
\quad $m_t$ &=& 172.74~GeV\,, &\quad $m^{\overline{\mbox{MS}}}_b(m_b^{\overline{\mbox{MS}}})$ &=& 4.18~GeV\,, \\
\quad $m_\tau$ &=& 1.77686~GeV\,,\,.&\quad  
%\label{eq:param1} 
\end{tabular}
\end{equation}
We have selected the  normal mass ordering class for the active neutrino inputs. The mass of the lightest neutrino is fixed to be $m_{\nu_1}=10^{-11}\; \gev$  while the other neutrino masses and mixing parameters are  chosen randomly from the $3\sigma$ ranges around their best-fit values,
\bea 
 m_{\nu_2} &\in&  [\sqrt{ m_{\nu_1}^2 + 6.82\times 10^{-23}},\sqrt{ m_{\nu_1}^2 + 8.04\times 10^{-23}} ]\, \gev\,,\crn
  m_{\nu_3} &\in&  [\sqrt{ m_{\nu_1}^2 + 2.435\times 10^{-21}},\sqrt{ m_{\nu_1}^2 + 2.598\times 10^{-21}} ]\, \gev\,,\crn
\theta_{12} &\in&  [\arcsin(\sqrt{0.269}),\arcsin(\sqrt{0.343})]\,,\crn
\theta_{23} &\in&  [\arcsin(\sqrt{0.415}),\arcsin(\sqrt{0.617})]\,,\crn
\theta_{13} &\in&  [\arcsin(\sqrt{0.02052}),\arcsin(\sqrt{0.02428})]\,,\crn
\delta_{CP} &\in&  [120,369]\,.
\eea

In order to study the impact of (s)neutrino contributions on the
effective triple Higgs coupling and on the Higgs-to-Higgs decays,
we select a valid parameter point from our scan sample and vary each neutrino and sneutrino
parameter individually while keeping the other parameters
fixed. Our chosen parameter point is
called {\tt P1}, with the following input parameters:
 \begin{align}
M_{H^\pm}            &= 1970   \, \gev \,,   & m_{\ti X}      & =\text{diag}(806,1204,305)\, \gev \,,       \crn
M_1             & =679  \, \gev  \,,      &m_{\ti N}          & =\text{diag}(337,157,359)\, \gev  \,,       \crn    
M_2              & =461  \, \gev   \,,   &A_{\nu}            & =\text{diag}(-678,-678,-678) \, \gev \,,       \crn             
\mueff              & =418 \, \gev    \,, &A_{X}              & =\text{diag}(-1019,-1019,-1019)\, \gev  \,,       \crn      
m_{\tilde{Q}_3} & = 1963 \, \gev \,,      &B_{\mu_X}          & =\text{diag}(31,31,31) \, \gev\,,       \label{eq:pointP1}\\
 m_{\tilde{t}_R}     & =2203 \, \gev \,,  &  M_X              & =\text{diag}(5179,7957,2527) \, \gev \,,       \crn            
m_{\tilde{\tau}_R}      &=647 \, \gev \,, & ((y_{\nu})_{11},(y_{\nu})_{12},(y_{\nu})_{13})
                                                     & =(-0.04,-0.48,-0.33)\,,       \crn
A_t                  & =-2982 \, \gev \,,  &    ((y_{\nu})_{21},(y_{\nu})_{22},(y_{\nu})_{23})
                                                     & =(-0.95,0.35,-0.04)\,,      \crn        
\mbox{Re} A_k             & =-2480 \, \gev   \,,   & ((y_{\nu})_{31},(y_{\nu})_{32},(y_{\nu})_{33})
                                                     &
                                                     =(-0.85,0.07,-0.87)\,,       \crn        
\tan\beta           & =6.4  \,,        &     
                                                                        \crn     
\lambda             &= 0.47   \,,      &             
\kappa              &=0.75  \,.        &  \nonumber
\end{align}
The resulting spectrum of the Higgs boson masses at ${\cal
  O}(\alpha_t(\alpha_s +\alpha_t))$ is given in
\tab{tab:HiggsSpectrumP1OS}. For this parameter point, the SM-like
Higgs boson is dominated by the $h_u$ component and its mass is about
125.0 GeV, while the second lightest Higgs boson has a mass of
$\unit[433.3]{GeV}$ and is dominantly $h_s$-like. The remaining Higgs
bosons are much heavier. 
 
\begin{table}[t!]
\begin{center}
\begin{tabular}{|c|c|c|c|c|c|c|c|}
\hline
                         &  ${H_1}$   
                              &  ${H_2}$  &  ${H_3}$  &  ${H_4}$  
                              &  ${H_5}$ 
                               \\ \hline \hline
ISS &125.0 & 433.3 & 1969.4  & 1971.6 &2197.0 \\
\hline
no-ISS & 123.1 & 433.3 &  1969.4 & 1971.7 & 2197.0
              \\ \hline
  &$ h_u$     & $h_s$   & $a$   & $h_d$   & $a_s$
  \\ \hline
\end{tabular}
\caption{The parameter
  point {\tt P1}:  Higgs  masses in GeV for the NMSSM
  with and without inverse seesaw mechanism and  their main components. 
  The mass spectrum was obtained at two-loop
    ${\cal O}(\alpha_t (\alpha_s + \alpha_t))$ with OS renormalisation in the
    top/stop sector.
}
\label{tab:HiggsSpectrumP1OS}
\end{center}
\vspace*{-0.4cm}
\end{table} 
%%%%%%%%%%%%%%%%%%%%
\subsection{Impact on the SM-like Effective Triple Higgs Coupling and the SM-like Higgs Mass}
%%%%%%%%%%%%%%%%%%%%
\begin{figure}[h]
\centering
\begin{tabular}{ccc}
	\includegraphics[height=7.2cm,width=7.2cm]{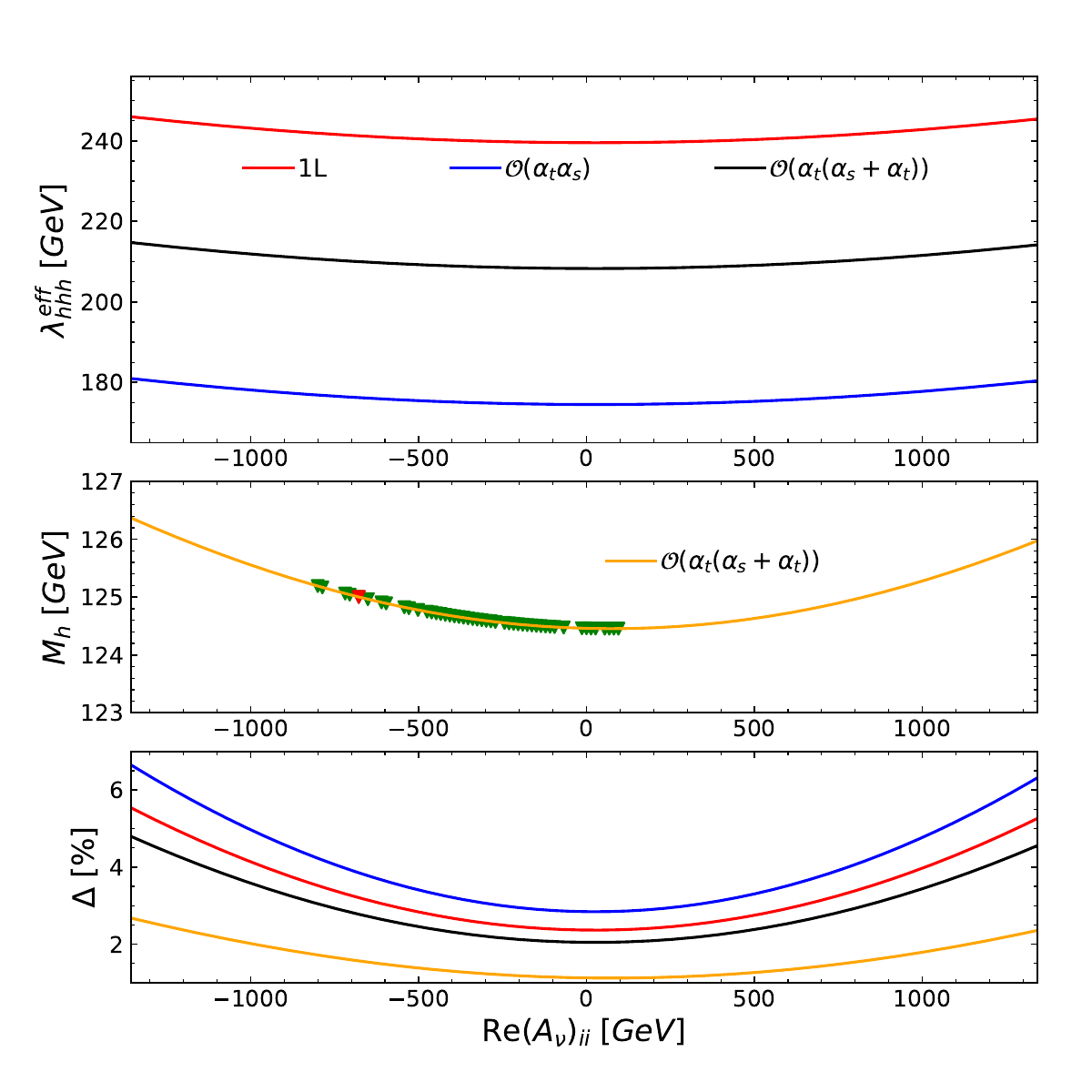} & 
	\includegraphics[height=7.2cm,width=7.2cm]{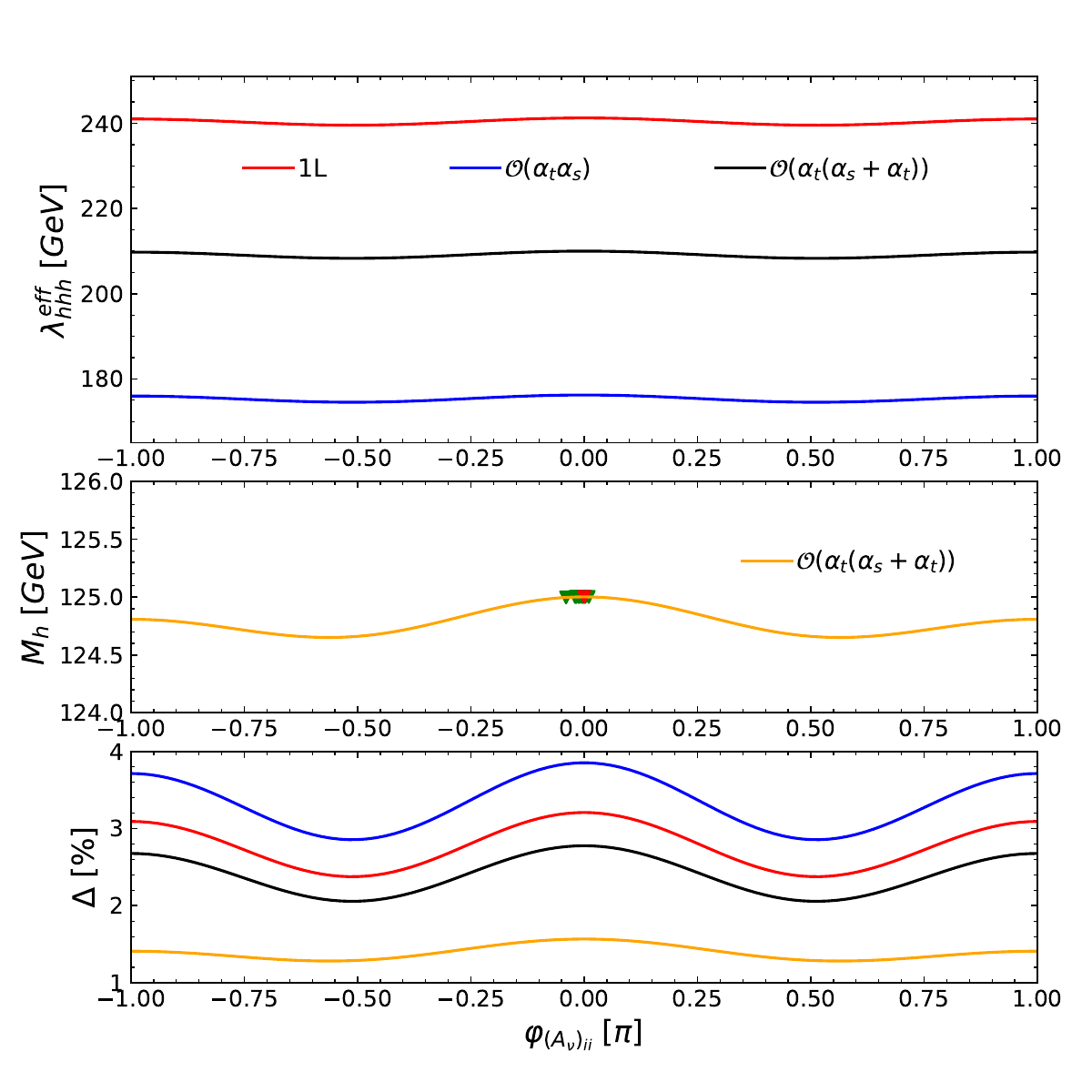}
\end{tabular}
\caption{Upper left panel: the effective trilinear coupling of the
  SM-like Higgs boson at one-loop order (red), 
$\oorder{\alpha_t \alpha_s}$ (blue) and  $\oorder{\alpha_t (\alpha_s+\alpha_t)}$ (black) as function of 
$\Re(A_\nu)_{ii}$. Middle left panel: the yellow line presents the
SM-like Higgs mass $\oorder{\alpha_t (\alpha_s+\alpha_t)}$ as function
of $\Re(A_\nu)_{ii}$, 
 the red triangle represents the parameter point P1, green triangles are points satisfying all constraints. Lower left 
 panel:  relative changes of the
   corrections defined in \eqref{eq:Delta} to the trilinear Higgs self-coupling at the various loop orders  (blue,
   red, black lines) and to the Higgs boson mass at $\oorder{\alpha_t
     (\alpha_s+\alpha_t)}$ (yellow line) (in percent) 
due to the (s)neutrino sector. Right plots: similar to left
 plots but  the phase of $(A_\nu)_{ii}$ is varied instead.
 }
\label{fig:Av11}
\end{figure}

We first  investigate the dependence of the effective trilinear
coupling $\lambda^{\eff}_{hhh}$ of the SM-like Higgs boson $h$ on a selection of SUSY input parameters. 
In order to do so, we take the parameter 
point {\tt P1} and vary the  chosen input
parameter around its central value in {\tt P1},
cf.~\eqref{eq:pointP1}. In \figref{fig:Av11}~(left), we
vary the real part of $(A_\nu)_{ii}$ ($i=1,2,3$)  while in \figref{fig:Av11}~(right)  the corresponding complex phases are varied. Note that we have chosen $A_\nu$ to be diagonal
 with
 the three diagonal entries taken to be identical. The upper panels in \figref{fig:Av11}
show the effective trilinear Higgs coupling of the SM-like
Higgs at one-loop order (red), $\oorder{\alpha_t \alpha_s}$ (blue) and  $
 \oorder{\alpha_t (\alpha_s+\alpha_t)}$ (black). We remind the reader that the 
 contribution from the (s)neutrino sector 
   only enters at the one-loop level,  whereas the two-loop 
 corrections are from the top/stop sector only and do not depend on
 the neutrino parameters. On the
  middle panels, we plot the predictions for the SM-like Higgs boson mass at 
  $\oorder{\alpha_t (\alpha_s+\alpha_t)}$ (yellow line) as function of the varied 
  parameters. The predictions for the originally chosen parameter point {\tt P1} are presented by the red 
  triangle while all parameter points satisfying our applied
  constraints are denoted by green triangles. In the lower panels,
  we quantify the contribution from the (s)neutrino sector by plotting the relative differences between the results in
\NMSSMCALCnuSS and in \NMSSMCALC, which is defined as 
\be \Delta= \fr{x^{\text{ISS}}-x^{\text{no-ISS}} }{x^{\text{no-ISS}}} , \label{eq:Delta} \ee
  with $x^{\text{ISS}}$ 
 ($x^{\text{no-ISS}}$) referring to 
 the SM-like trilinear self-couplings or the Higgs boson mass in the NMSSM 
 with (without) inverse seesaw mechanism at the same
 loop order.  The resulting figure suggests that the trilinear and the Higgs mass values are symmetric around
 $(A_\nu)_{ii}=0$, which is, however, not true for the following reasons. 
The minimal Higgs mass values are  obtained
for  $\Re(A_\nu)_{ii} = \Re(\mueff)/\tbeta\sim 65\, \gev$ for {\tt P1},
which corresponds to the off-diagonal components  of the  
sneutrino mass matrix, $M_{\ti \nu_\pm \ti N_\pm}$, being zero. This
is similar to the dependence of the one-loop top/stop sector
contributions to Higgs boson masses on $A_t$ \cite{Muhlleitner:2014vsa,Dao:2019qaz}. For the trilinear Higgs self-couplings,
however, the dependence on $A_\nu$ is more complicated, and the
minimal values are obtained for $\Re(A_\nu)_{ii} \sim 30 \,\gev$. 
We observe that there are more valid points  for negative values of $\Re(A_\nu)_{ii}$ and $\varphi_{(A_\nu)_{ii}}$ 
than for positive ones. 
The invalid points violate either the
constraints on the charged lepton flavor-violating decays or 
on the oblique parameters $S,T,U$.  A general feature, as observed in
the lower panels of \figref{fig:Av11}, is that  
the relative difference in the corrections due to the
(s)neutrino sector  for the 
trilinear Higgs self-coupling are two to three times larger  (depending on
the loop-order) than for the SM-like Higgs mass. In particular, 
the relative difference in the corrections for the Higgs mass range 
between 1\% and 2.8\% for the varied range of $\Re(A_\nu)_{ii}$, while 
the ones for the trilinear couplings are in [3\%,5.8\%] at one-loop order, in [3.5\%,6.3\%] at  $\oorder{\alpha_t \alpha_s}$ and in [2.6\%,4.5\%] at
$\oorder{\alpha_t (\alpha_s+\alpha_t)}$. These
relative differences in the corrections as defined above give the impression of a rather
  moderate overall importance of the higher-order corrections to the trilinear
coupling. However, this is deceptive because we divide the (s)neutrino corrections by the 
 result obtained in \NMSSMCALC at the same considered
 order thereby accidentally canceling all higher-order
   contributions from the s/top sector
   of the denominator with the (s)neutrino
   contributions in the numerator.
 If we instead define the (s)neutrino corrections as 
\beq
 \delta= (\lambda_{hhh}^{\eff,\text{ISS},1L} -\lambda_{hhh}^{\eff,\text{no-ISS},1L})/\lambda_{hhh}^{\eff,\text{ISS},\text{Tree}}\,,
 \eeq
then we are only sensitive to the higher-order corrections originating
from the (s)neutrino sector and find that the one-loop (s)neutrino corrections are in the range of
 $\delta\propto [7\%,15\%]$.
These large corrections arise from several large neutrino
Yukawa coupling components such as
$(y_\nu)_{21},(y_\nu)_{31},(y_\nu)_{33}$ which are close to
one. \s 

We also observe a slight dependence of the trilinear Higgs
self-coupling and the SM-like Higgs mass on the phase of
$(A_\nu)_{ii}$, as can be inferred from the right panels of
\figref{fig:Av11}. We vary the phase in the range $[-\pi,\pi]$.
From the distribution of the green points in the middle right panel it can be seen that the phase $\varphi_{(A_\nu)_{ii}}$ is constrained to very small values. We note that the
electric dipole moment (EDM) measurements, which are known to tightly constrain the CP-violating phases in the NMSSM without seesaw mechanism, are not the dominating constraint in this scenario. Most of the points with larger phases pass EDM contraints but violate
the $S,T,U$ constraints and/or the constraints from the charged lepton
flavor-violating decays. \s

\begin{figure}[ht!]
\centering
\begin{tabular}{ccc}
	\includegraphics[height=9.2cm,width=9.2cm]{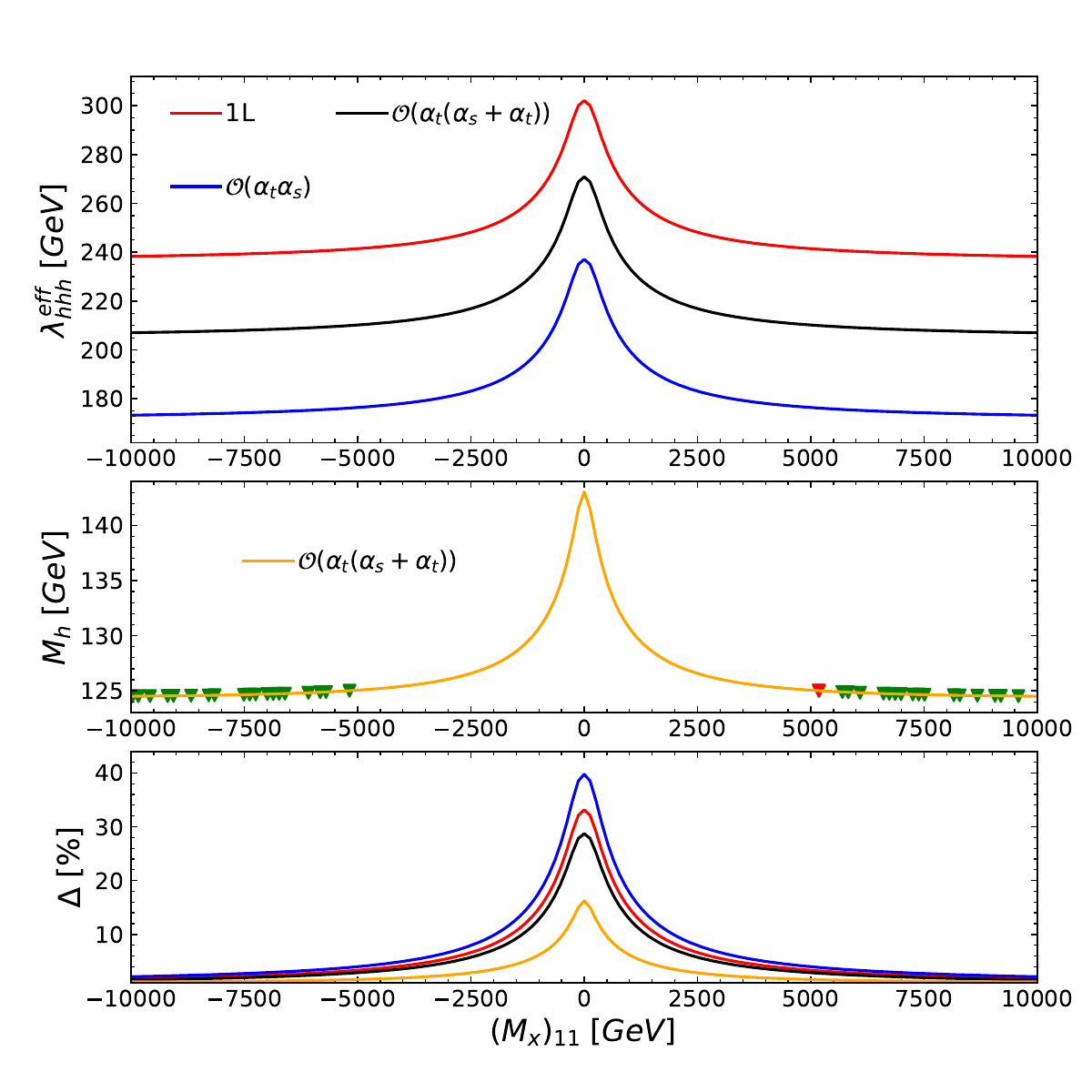} & 
\end{tabular}
\caption{Similar to \figref{fig:Av11} but $(M_X)_{11}$ is varied
  instead.}
\label{fig:MX11}
\end{figure}
We next show the impact of $M_X$ on the effective trilinear Higgs
self-couplings  $\lambda^{\eff}_{hhh}$. In Fig.~\ref{fig:MX11}, 
we present the results with a varying first
diagonal component of $M_X$. The dependence of 
$\lambda^{\eff}_{hhh}$ and $M_h$ on the other diagonal components
is qualitatively the same compared to the one shown here. The notation of Fig.~\ref{fig:MX11}
is the same as the one in  \figref{fig:Av11}.
The prominent feature of these plots is the
peak at $(M_X)_{11}=0$ in both the mass and the
trilinear Higgs coupling plot. While the mass can go up to 143 GeV,
the trilinear Higgs coupling
at $\oorder{\alpha_t (\alpha_s+\alpha_t)}$ goes up to 271
GeV.\footnote{We remind the reader that the tree-level SM value is
  given by
  $\lambda_{H_{\text{SM}}H_{\text{SM}}H_{\text{SM}}}=3M_{H_{\text{SM}}}^2/v\approx
  190$~GeV.} The dominant contributions to the corrections on the mass
as well on the trilinear Higgs couplings are from the
the two lightest sneutrinos and the two lightest heavy neutrinos whose main components are from the singlet superfield $\hat X_1$. 
For this parameter point, the masses of the two lightest sneutrinos  are around 378
GeV which are the smallest values for all considered values of $(M_X)_{11}$. They are, however, still not the lightest SUSY particle which is the lightest neutralino
 with a mass of 368 GeV.  The two lightest heavy neutrinos have a mass
 of 216 GeV. As usual, there is a cancellation between the contributions from fermions and their
 superpartners. At $M_X$, the fermions 
 contribute with a positive sign to the scalar diagrams while
 their superpartners contribute with a negative sign. The neutrino Yukawa
 coupling component $(y_\nu)_{21}=0.95$ enlarges these
 contributions. \s

\begin{figure}[t]
\centering
\begin{tabular}{ccc}
	\includegraphics[height=5.8cm,width=5.2cm]{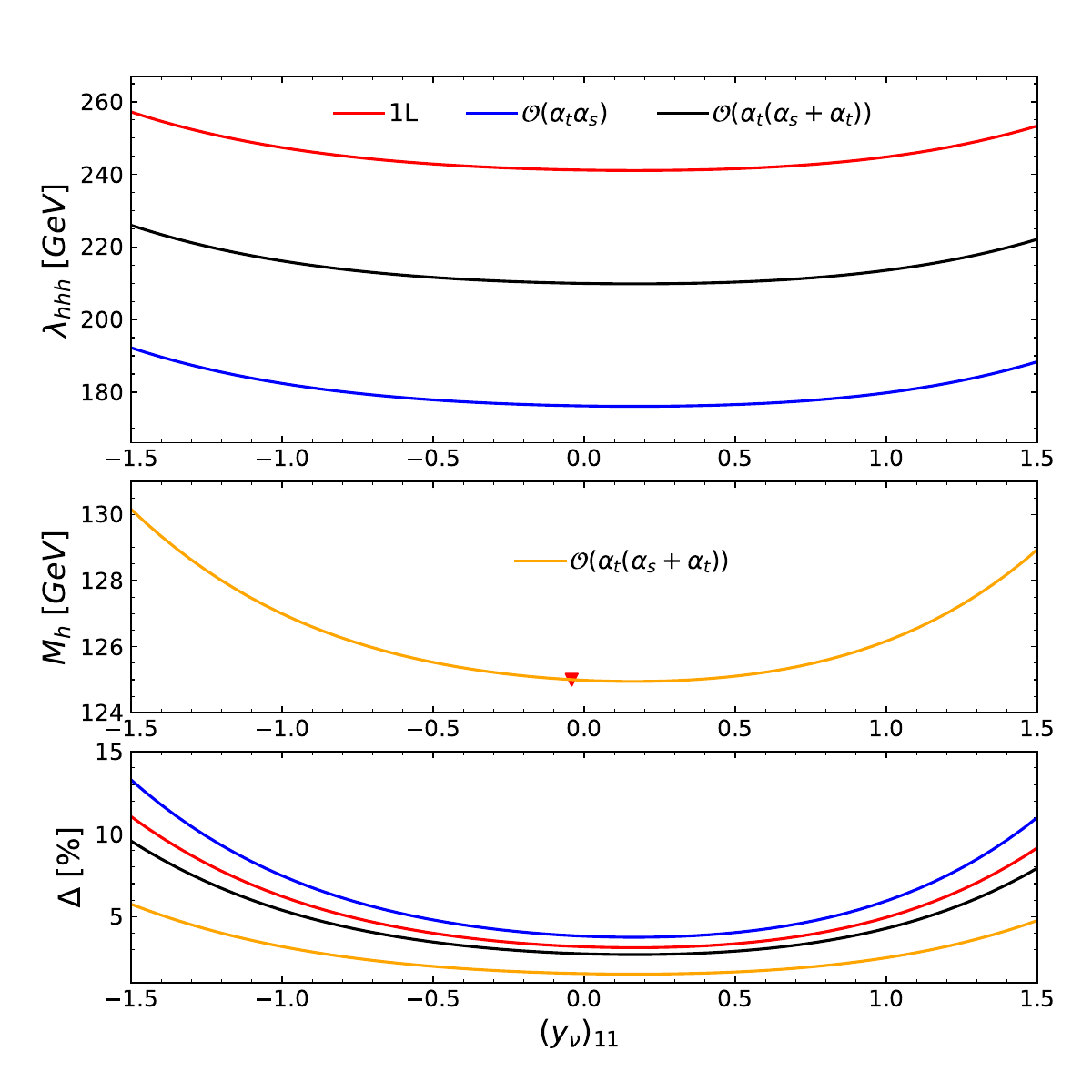} & 
	\includegraphics[height=5.8cm,width=5.2cm]{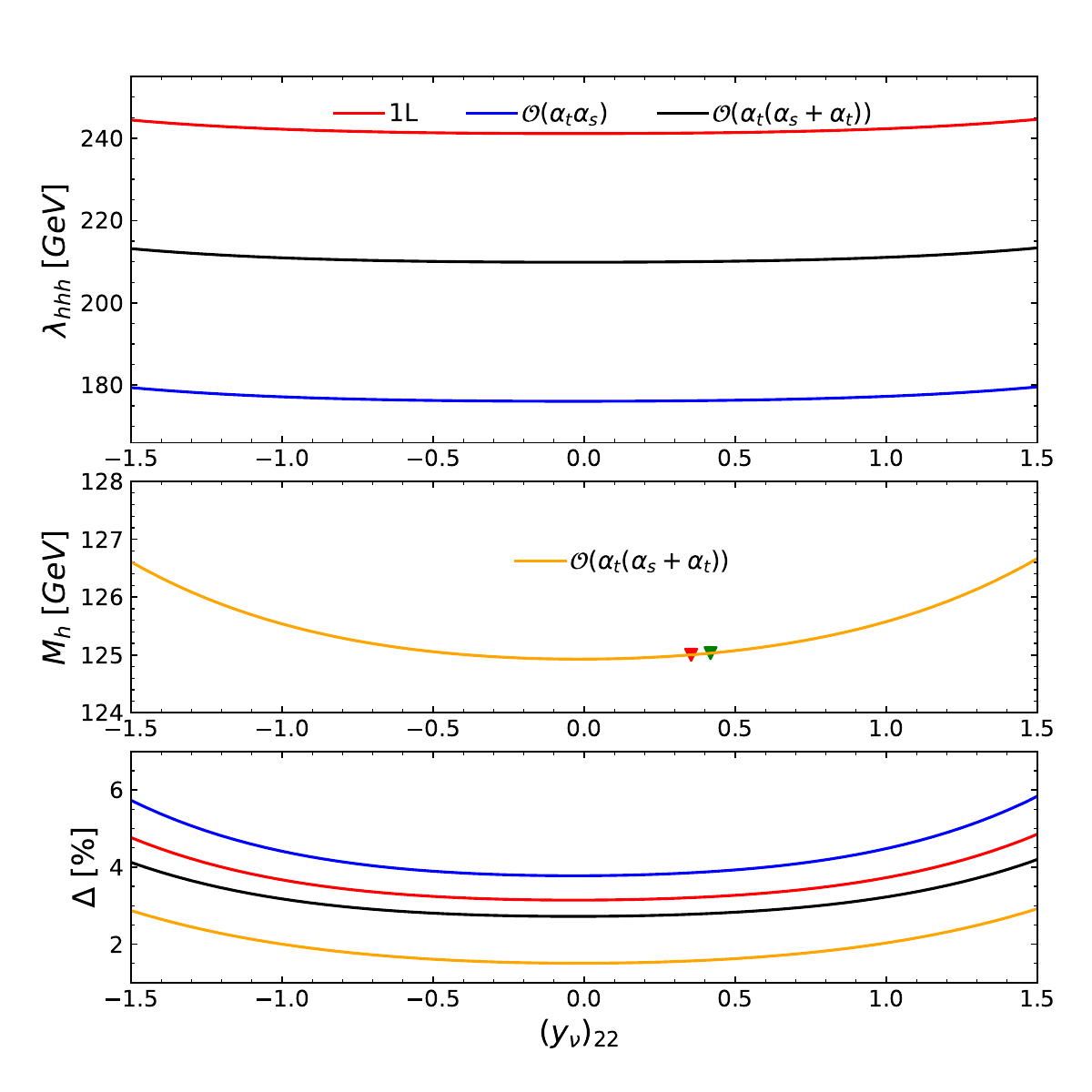} & 
	\includegraphics[height=5.8cm,width=5.2cm]{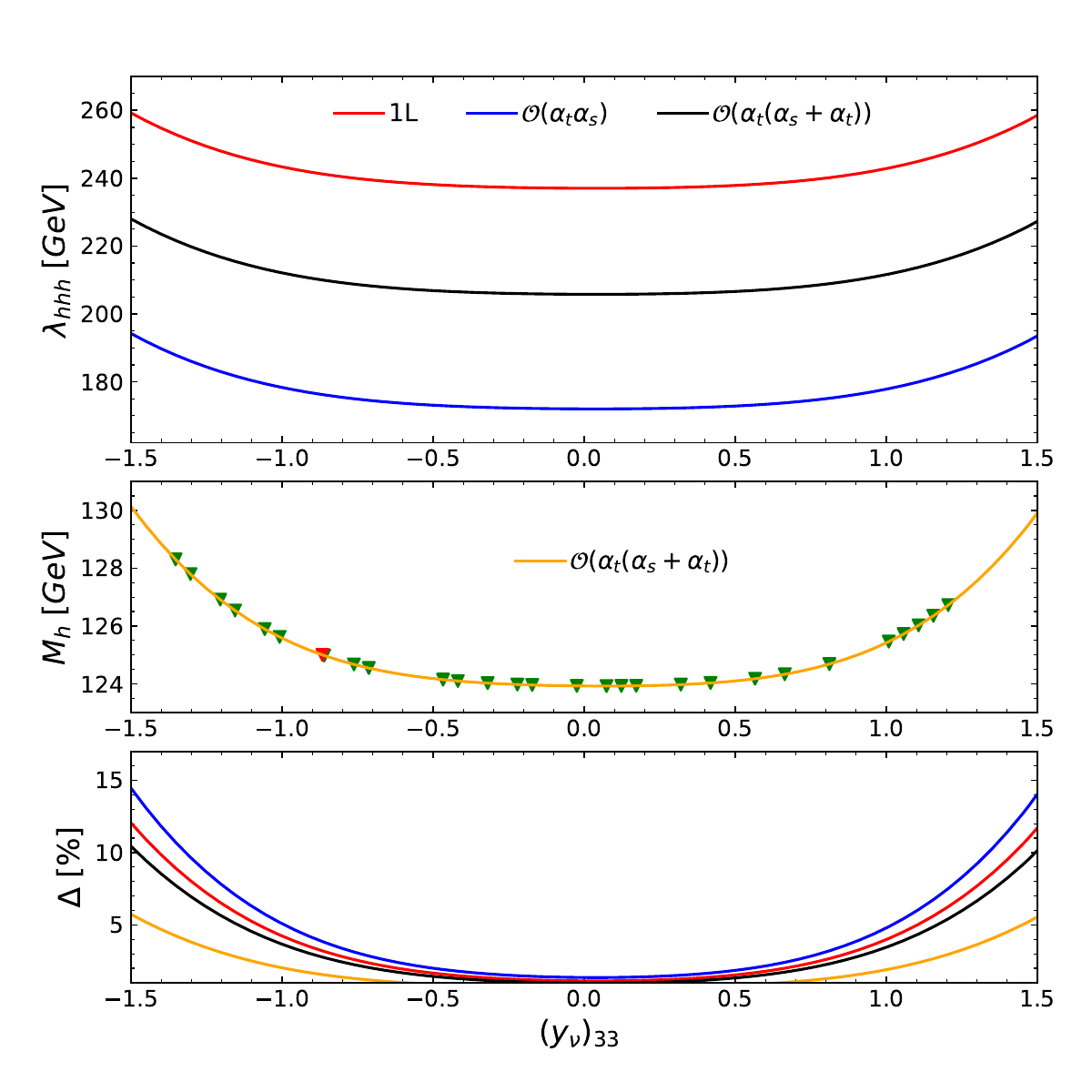} \\ 
	\includegraphics[height=5.8cm,width=5.2cm]{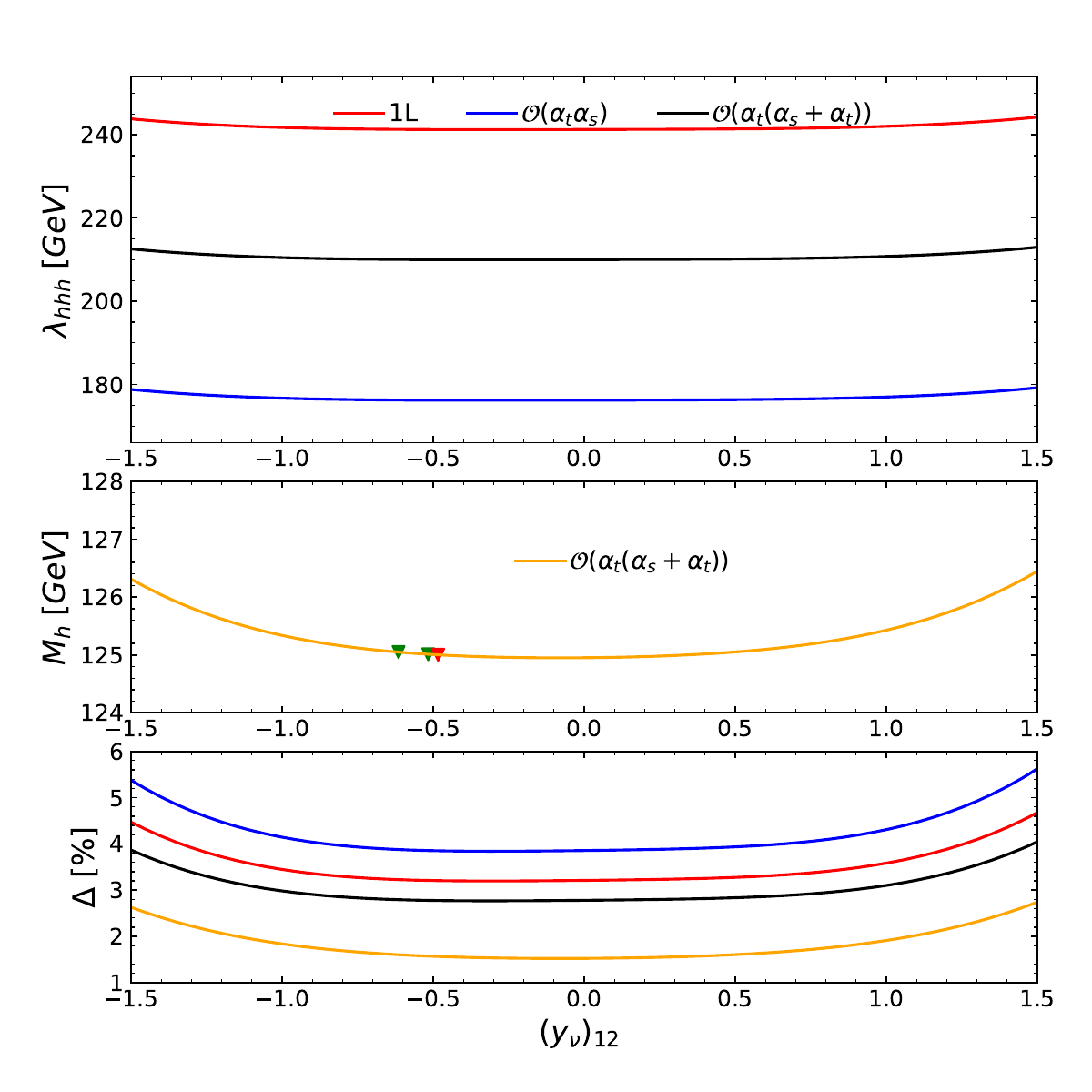} & 
	\includegraphics[height=5.8cm,width=5.2cm]{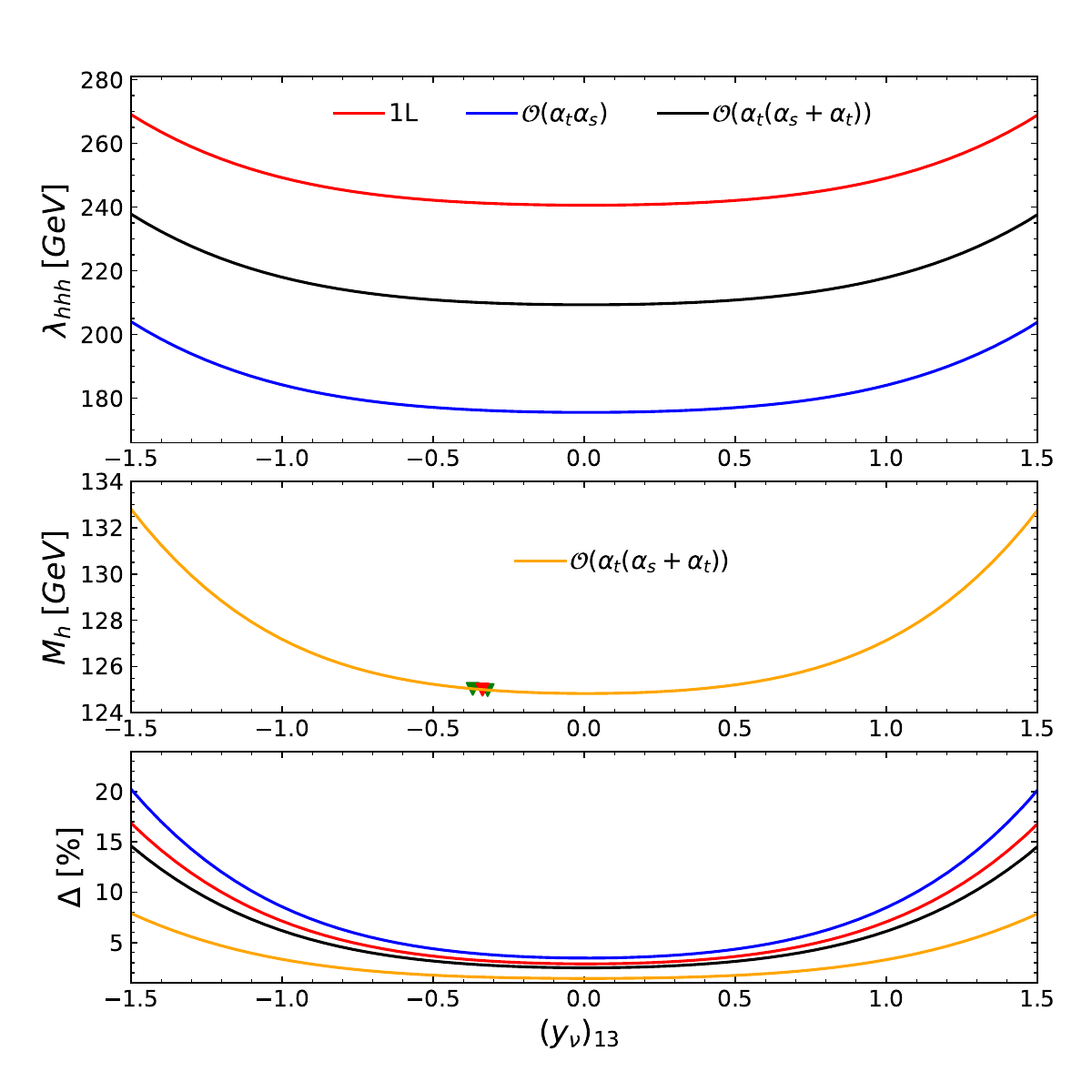} & 
	\includegraphics[height=5.8cm,width=5.2cm]{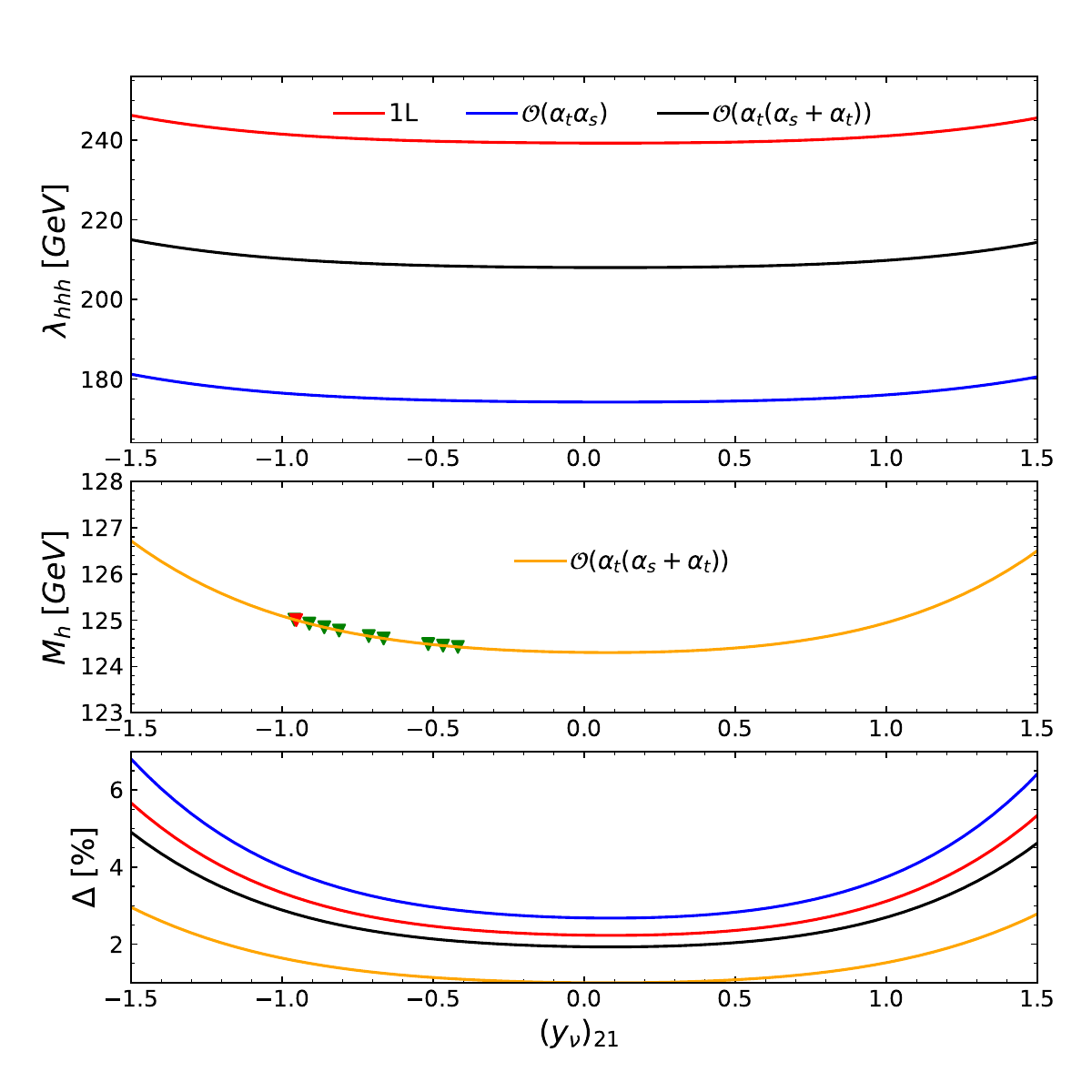} \\ 
	\includegraphics[height=5.8cm,width=5.2cm]{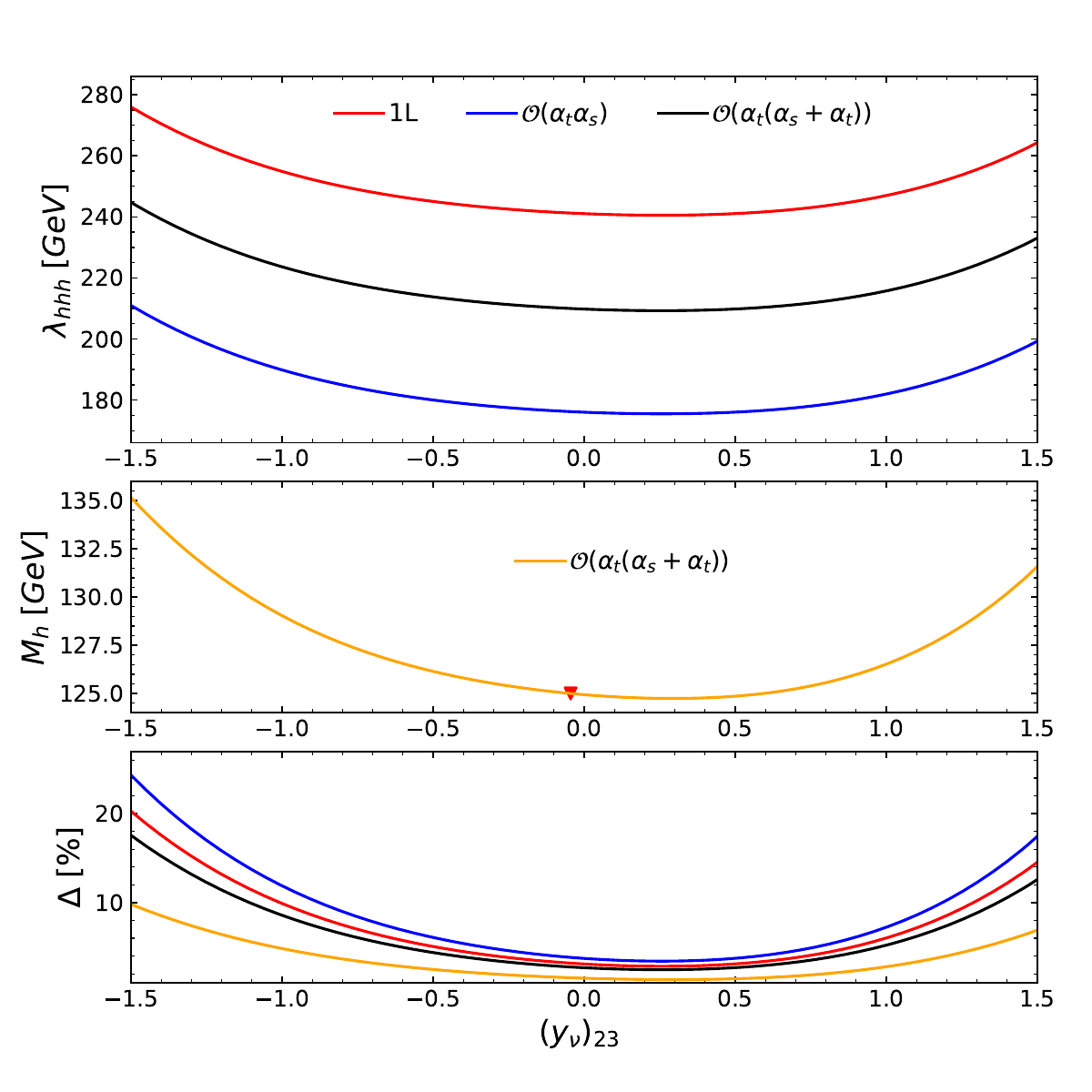} & 
	\includegraphics[height=5.8cm,width=5.2cm]{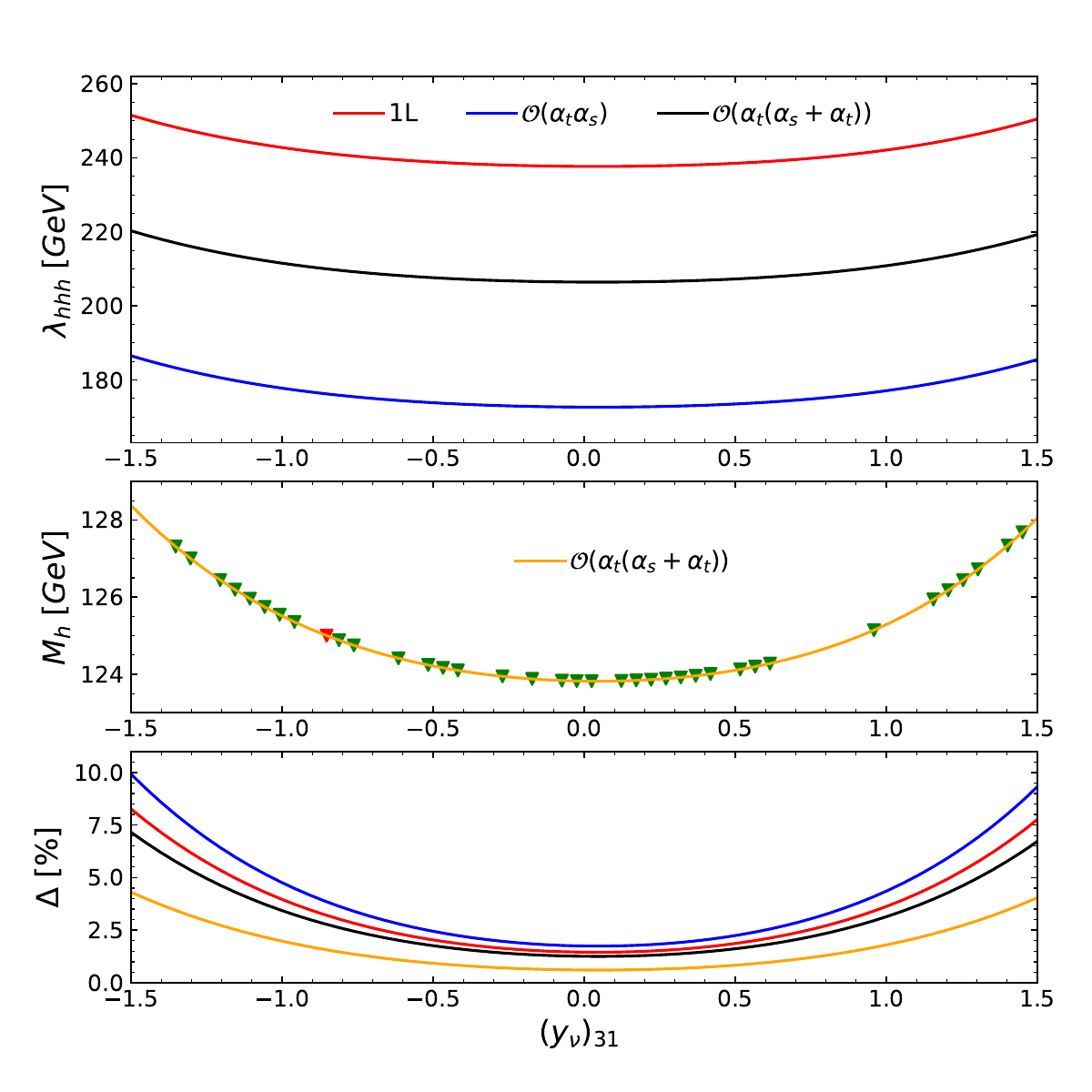} & 
	\includegraphics[height=5.8cm,width=5.2cm]{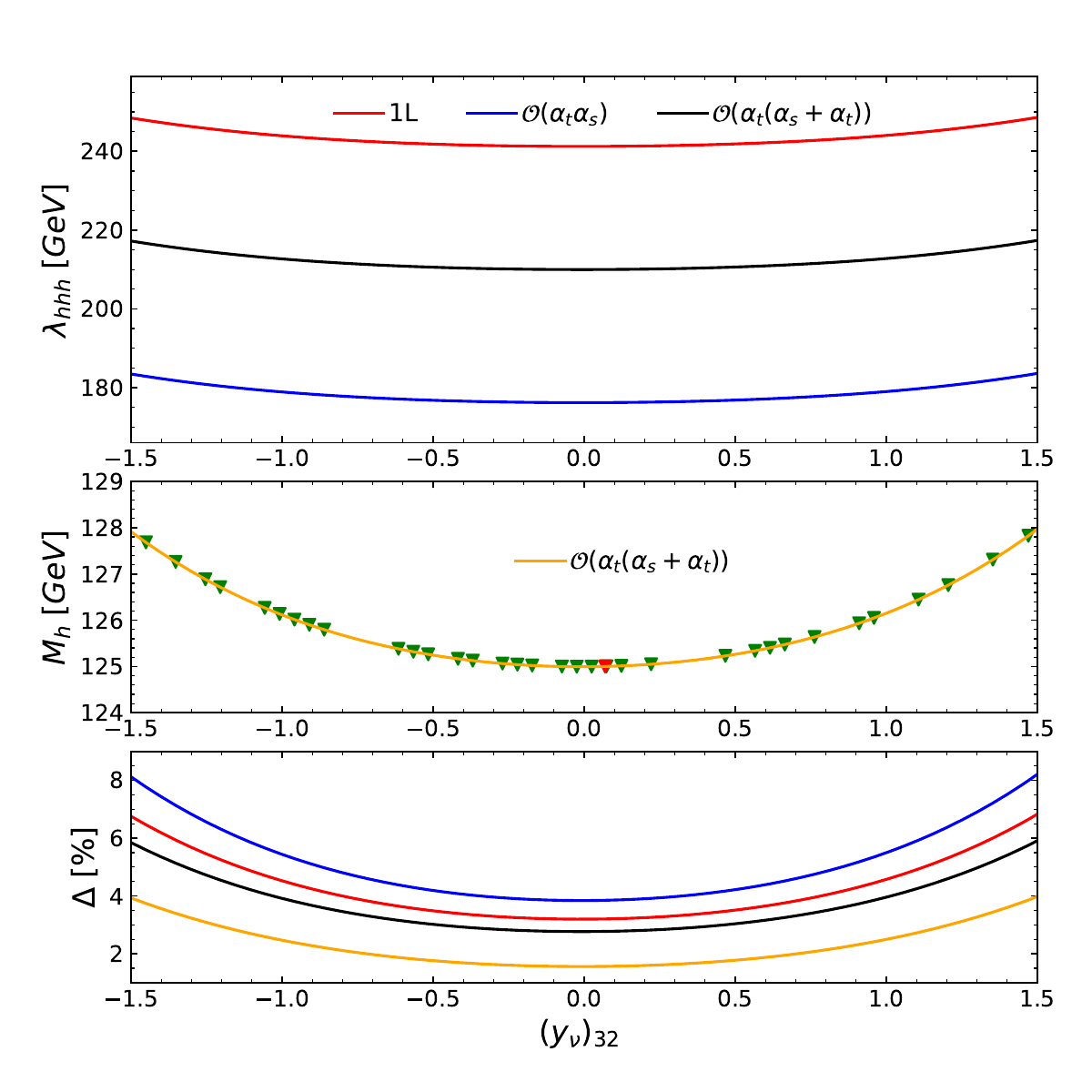} 
\end{tabular}
\caption{ Dependence of the SM-like triple Higgs self-coupling, the
  Higgs mass and the relative differences $\Delta$ on the 
nine components of the neutrino Yukawa matrix. The
notation and color code is the same as in Fig.~\ref{fig:Av11}.}
\label{fig:Yv}
\end{figure}

From the previous discussion we know that the neutrino Yukawa coupling
matrix $y_\nu$ is an important factor in the (s)neutrino contributions.
In \figref{fig:Yv} we present nine plots
in which the nine elements of the matrix $y_\nu$
are varied independently. The notation and color code
are the same as in the previous plots. 
We chose the ranges of all varied elements to be $[-1.5,1.5]$  which
is still much smaller than the perturbative unitary limit of
$\propto\sqrt{4\pi}$.  We observe that the different
elements of the matrix  affect the  
trilinear Higgs self-coupling and the mass  as
well other observables with a different
magnitude and shape. For
some elements like $(y_\nu)_{11}, (y_\nu)_{33}, (y_\nu)_{13},
(y_\nu)_{23}$, the quantity $\Delta$ 
for  $\lambda^{\eff}_{hhh}$ at $\oorder{\alpha_t (\alpha_s+\alpha_t)}$
can be larger than 10\%, while for other elements, it remains below  
7.5\%. \s

Furthermore, we do not observe a symmetric behaviour around
$y_{\nu}=0$. We also observe that the constraints from 
the $S,T,U$ parameters and the charged lepton flavor-violating decays
are very sensitive to the change of each element. In particuler, we still get
a large number of valid points when
varying the third row elements $(y_\nu)_{31}, (y_\nu)_{32}, (y_\nu)_{33}$,
which is
not the case for the other elements that are
in-turn much more strongly constrained.\s

\begin{figure}[h]
\centering
\begin{tabular}{ccc}
	\includegraphics[height=6.8cm,width=7.2cm]{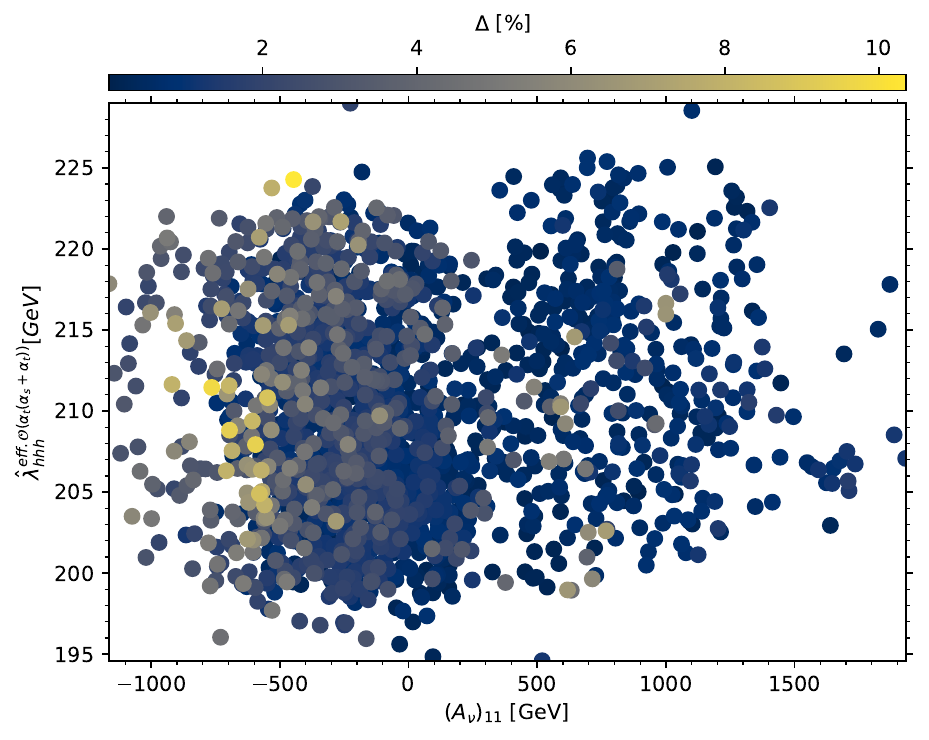} & 
	\includegraphics[height=6.8cm,width=7.2cm]{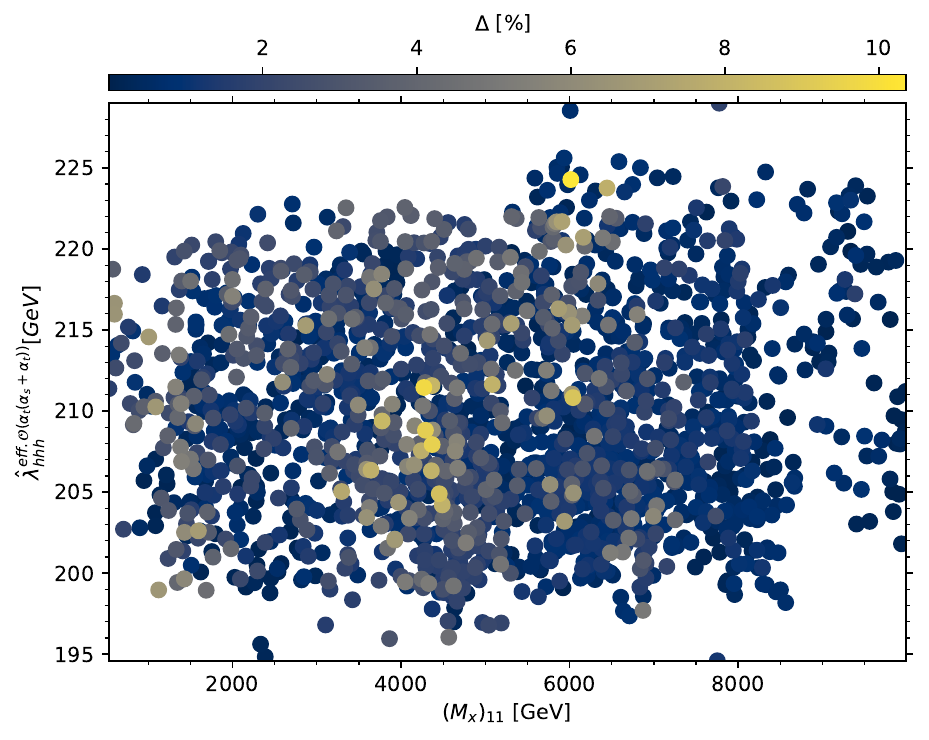} \\
	\includegraphics[height=6.8cm,width=7.2cm]{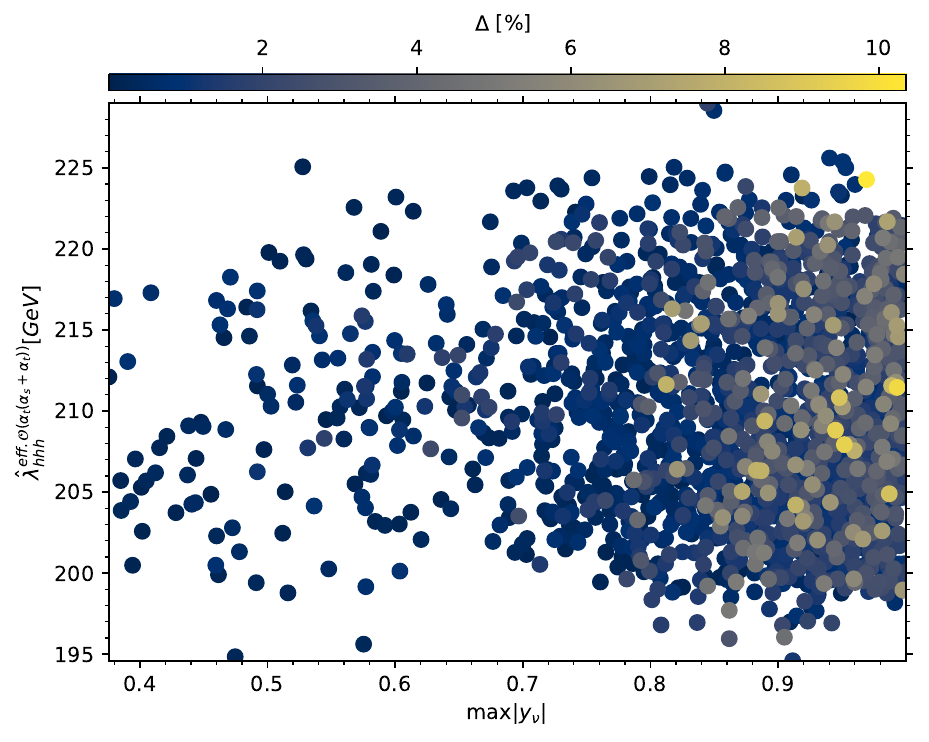} & 
	\includegraphics[height=6.8cm,width=7.2cm]{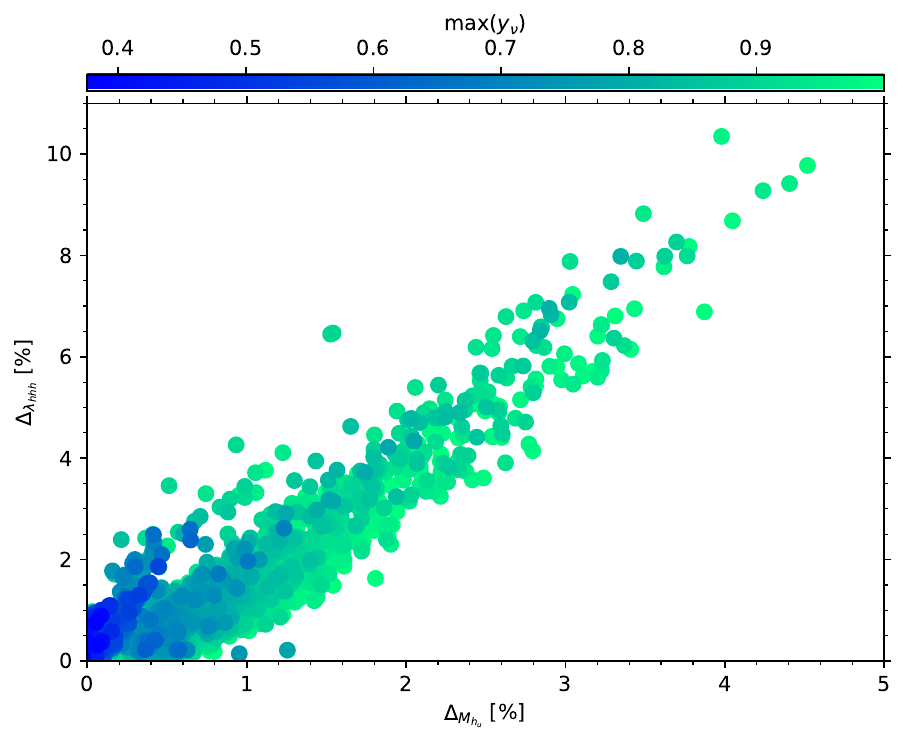}  
\end{tabular}
\caption{Scatter plots for the effective SM-like triple Higgs
  self-couplings (upper and lower left)  at $\oorder{\alpha_t 
    (\alpha_s+\alpha_t)}$ for all parameter points obtained from our
  scan satisfying all our constraints.  The color code of each point
  indicates the relative (s)neutrino correction $\Delta$ in percent. 
  The lower right plot shows the correlation between the relative (s)neutrino corrections for the  effective SM-like triple Higgs
  self-coupling and for the SM-like Higgs mass. The color code indicates the maximal absolute  value of all neutrino Yukawa coupling elements at each point. 
}
\label{fig:scatter1}
\end{figure}

We close this section by showing in \figref{fig:scatter1} scatter plots
obtained in the parameter scan described at the
beginning of Sec.~\ref{sec:num_anal}. In the upper left (upper right / lower
left) plot we show the effective trilinear Higgs self-couplings computed at
the highest available order as a function of $(A_\nu)_{11}$ ($(M_X)_{11}$ /
max$|y_\nu|$), where max$|y_\nu|$ refers to the component of $y_\nu$ with the
largest absolute value. All parameter points shown in the plots satisfy 
the applied constraints discussed above. The color of each point represents the magnitude of the 
relative differences $\Delta$ in percent. For all valid points
we find that the trilinear Higgs self-coupling $\lambda^{\eff}_{hhh}$ at $\oorder{\alpha_t
(\alpha_s+\alpha_t)}$ is
in the range of [195,229] GeV. From these plots, one can infer the
preferred regions in the parameters space to 
get both valid points and maximal (s)neutrino contributions. 
In the lower right panel, we show a scatter plot with 
the relative (s)neutrino corrections computed at $\oorder{\alpha_t
  (\alpha_s+\alpha_t)}$ for the  effective SM-like triple Higgs 
self-coupling on the vertical axis and the one for the SM-like Higgs
mass on the horizontal axis. We observe a strong correlation 
between these two corrections, similar to the findings in
\cite{Borschensky:2022pfc}, showing that the simultaneous inclusion of
higher-order corrections to both quantities can be equally important.
For the found parameter points, the values of $\Delta$ can go up to 10.5\%
 for the trilinear Higgs self-coupling and up to 4.5\% for the Higgs mass,
 respectively.
    
%%%%%%%%%%%%%%%%%%%%%%%%%%%%%%%%%%%%%%%%%%%%%%%%%%%%%%%%%%%%%
\subsection{Impacts on Higgs-to-Higgs Decays}
In this section we present our new results for the
Higgs-to-Higgs decay widths and branching ratios
implemented in the code \NMSSMCALCnuSS.  We first investigate the parameter point {\tt P1},
cf.~\tab{tab:HiggsSpectrumP1OS}.
In \tab{tab:Higgsdecays} we show all kinematically allowed Higgs decay
channels with the predictions for their
corresponding branching ratios at tree-level, one-loop order,
$\oorder{\alpha_t \alpha_s}$ 
and $\oorder{\alpha_t (\alpha_s+\alpha_t)}$ in the NMSSM with inverse
seesaw mechanism. For comparison, we present in the last
row results for the branching ratios at $\oorder{\alpha_t
(\alpha_s+\alpha_t)}$ obtained in the NMSSM without inverse seesaw mechanism. Note that the  
Higgs-to-Higgs decay widths are defined in \eqref{eq:DCwidths} and that
we dress \textit{all} results with the external-leg corrections in order to
ensure the fields are on the mass shell. This means in particular,
that, in the defined tree-level accuracy, we include partially higher-order
effects through the WFR factor ${\bold Z}^H$. As we can see, there is a significant change of the
branching ratios when we compare the tree-level to the one-loop
results. Especially for the decay channel $H_2\to H_1 H_1$, the
one-loop result is about 40\% larger than the 
tree-level one. However, for the remaining decay channels, the
branching ratios are reduced when the higher-order
  corrections to the
decay widths are included. The effect of the (s)neutrino contributions
on the branching ratios
can be roughly estimated by comparing the 
branching ratios between the models with and without ISS at the same order of
accuracy. The contributions  entering the decay widths can be
classified into three parts: the Higgs masses,
the WFR factor and the one-loop diagrams. Since the effect from the
(s)neutrino sector is mostly related to the neutrino Yukawa couplings,
which are the couplings between the doublet 
Higgs $H_u$ and the left-handed neutrinos and
$N$-neutrinos, one expects that
their effect may be small for the decays of  the non-SM-like Higgs bosons.
 From the two last rows in  \tab{tab:Higgsdecays}, one can see that 
the new contributions are indeed small on the branching
ratios. The
relative differences $\Delta$ are largest for the decay $H_4\to H_1H_1$
with about $3.4\%$. \s

\begin{table}[t!]
\begin{center}
\begin{tabular}{|c|c|c|c|c|c|c|c|}
\hline
                  &         \scriptsize{$\text{Br}({H_2\to H_1H_1})$   }
                              &  \scriptsize{$\text{Br}({H_4\to H_1H_1})$}  & \scriptsize{ $\text{Br}(H_4\to H_1H_2)$}  & \scriptsize{ $\text{Br}(H_4\to H_2H_2)$}
                              &  \scriptsize{ $\text{Br}({H_5\to H_1H_3})$}   
                               \\ \hline \hline
\scriptsize{Tree-level-ISS}&26.35 & 0.247 & 11.39 &   0.845 &  0.335 \\
\hline
\scriptsize{One-loop-ISS}&37.52 & 0.177 & 10.47 & 0.801& 0.327 \\
   \hline
   \scriptsize{$\oorder{\alpha_t \alpha_s}$-ISS} &35.11 & 0.206 & 10.45 & 0.804& 0.323 \\
   \hline
  \scriptsize{$\oorder{\alpha_t (\alpha_s+\alpha_t)}$-ISS} & 36.46 & 0.215 & 10.44 &   0.804 &  0.323 \\
   \hline
  \scriptsize{$\oorder{\alpha_t (\alpha_s+\alpha_t)}$-no-ISS} & 36.03 & 0.208 & 10.38 &   0.788 &  0.321 \\
   \hline   
\end{tabular}
\caption{The parameter
  point {\tt P1}: all possible Higgs-to-Higgs decays and their branching
  ratios in percent at several levels of accuracy in the
  NMSSM with and without inverse seesaw mechanism.
}
\label{tab:Higgsdecays}
\end{center}
\vspace*{-0.4cm}
\end{table} 

\begin{figure}[t]
\centering
\begin{tabular}{ccc}
	\includegraphics[height=5.8cm,width=5.2cm]{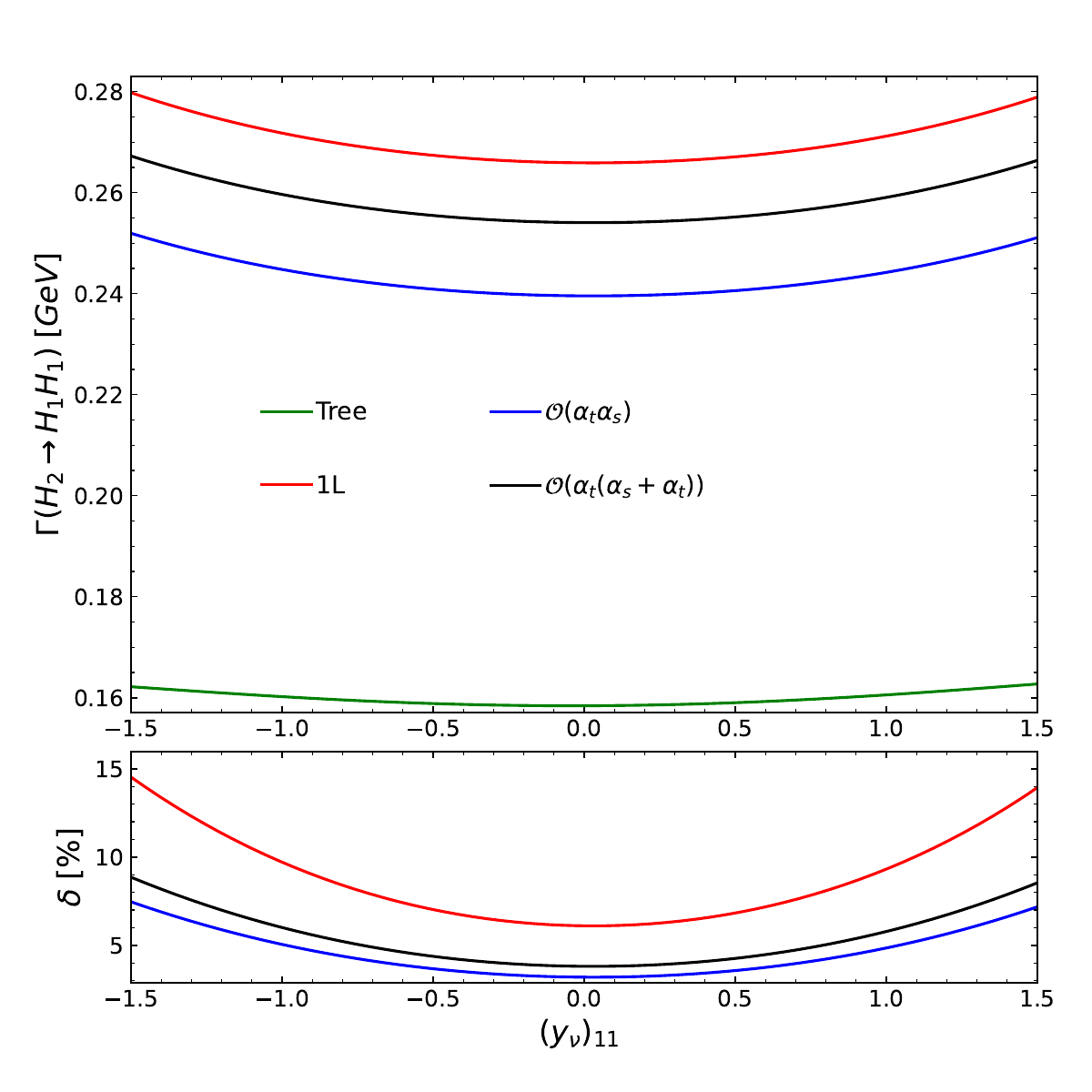} & 
	\includegraphics[height=5.8cm,width=5.2cm]{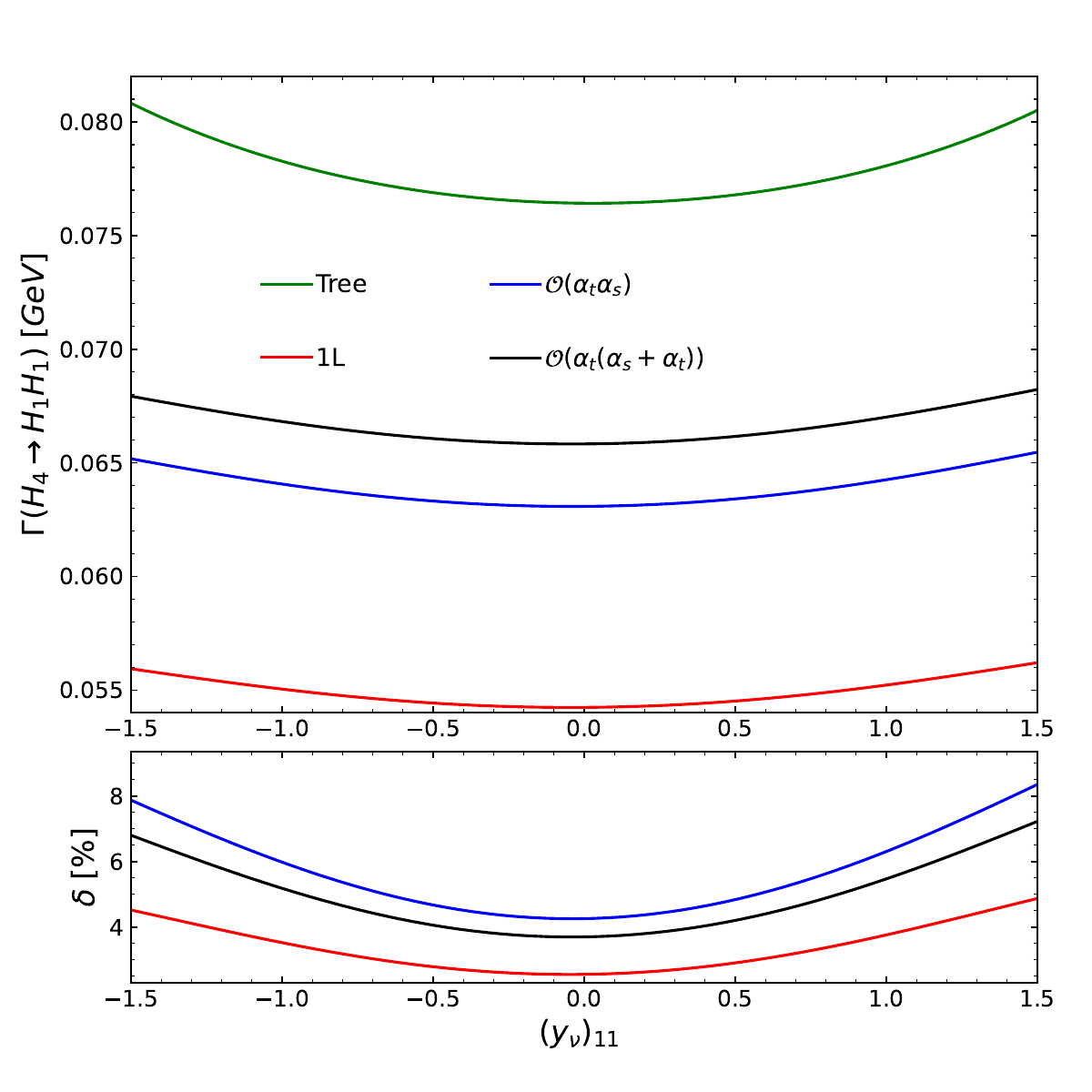} & 
	\includegraphics[height=5.8cm,width=5.2cm]{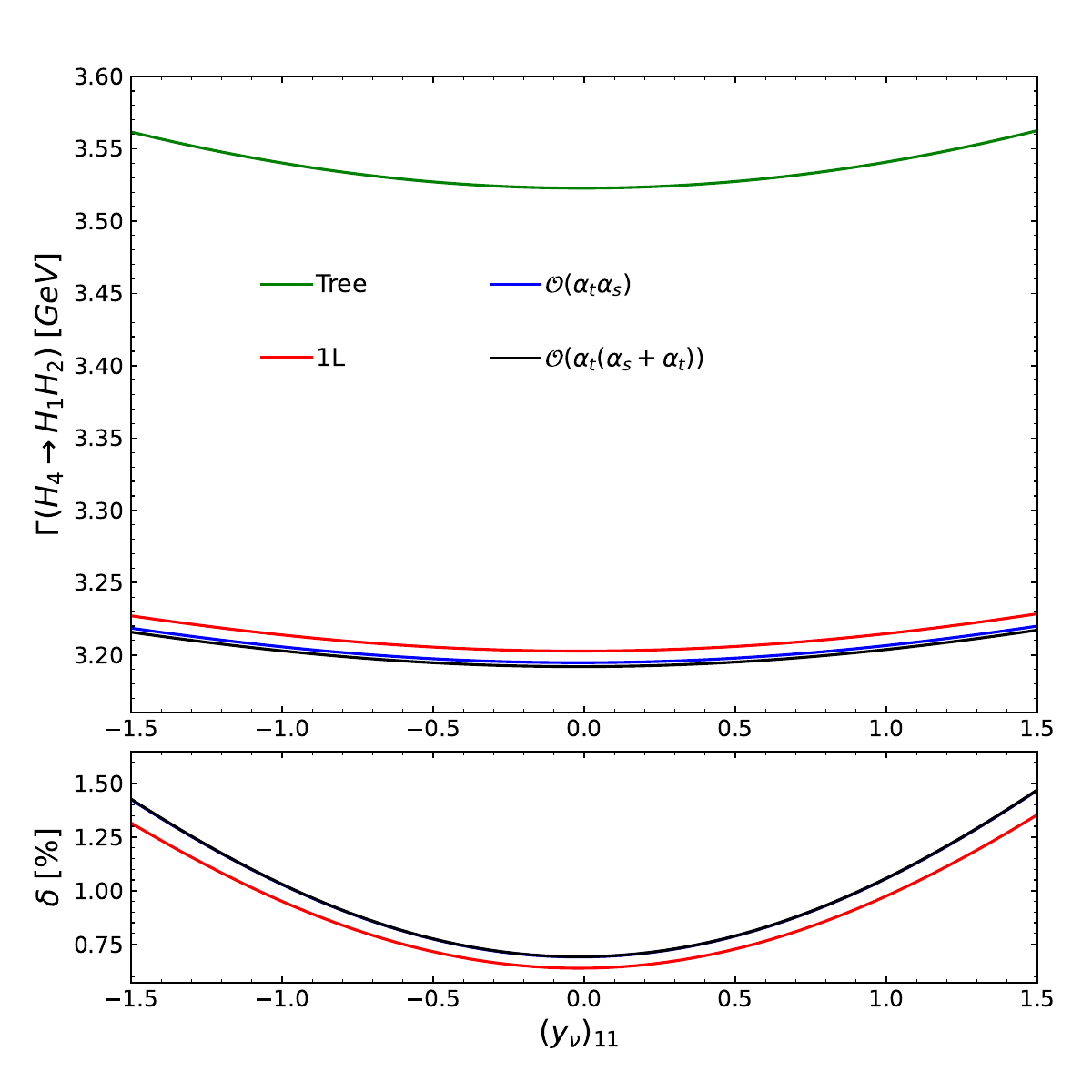} \\ 
	\includegraphics[height=5.8cm,width=5.2cm]{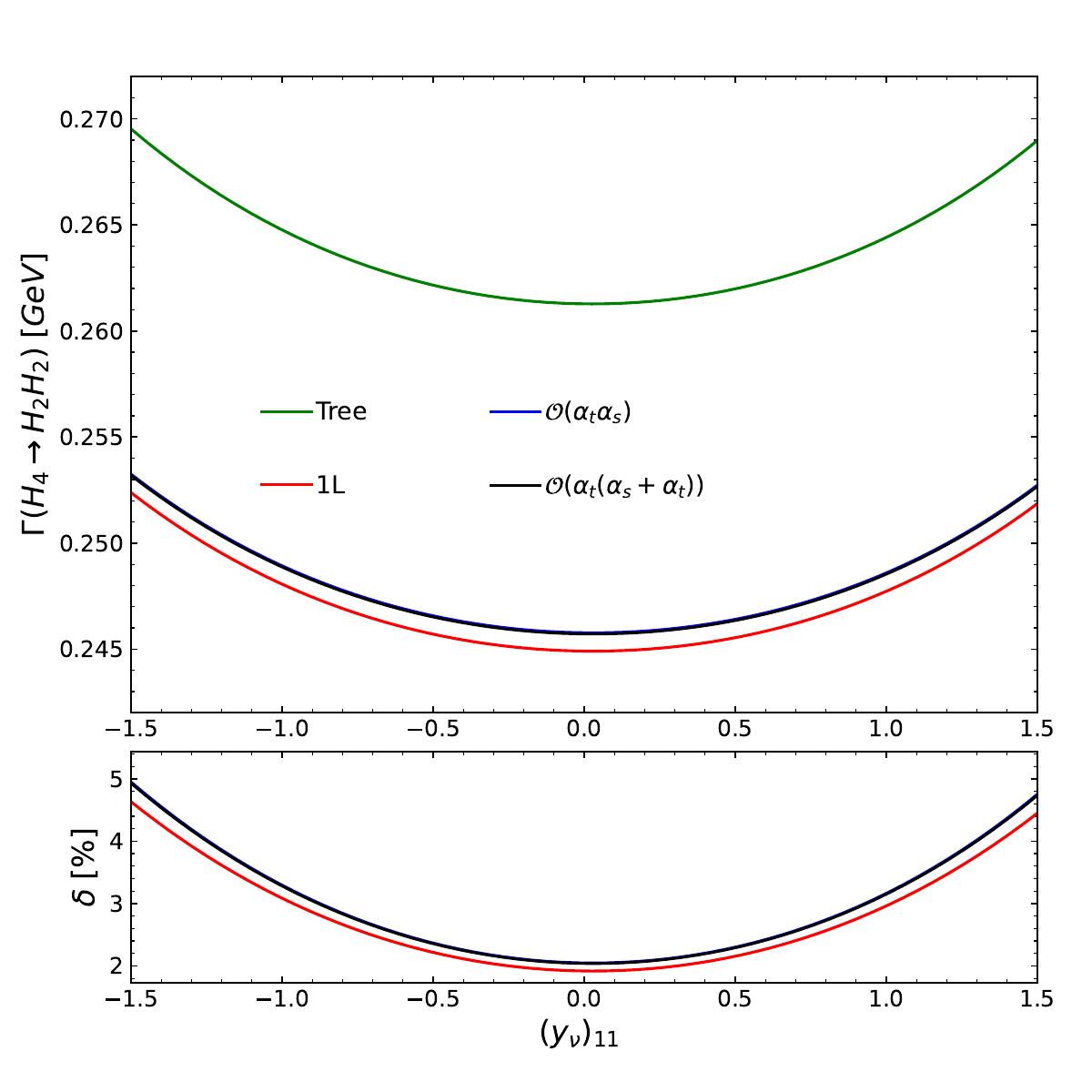} & 
	\includegraphics[height=5.8cm,width=5.2cm]{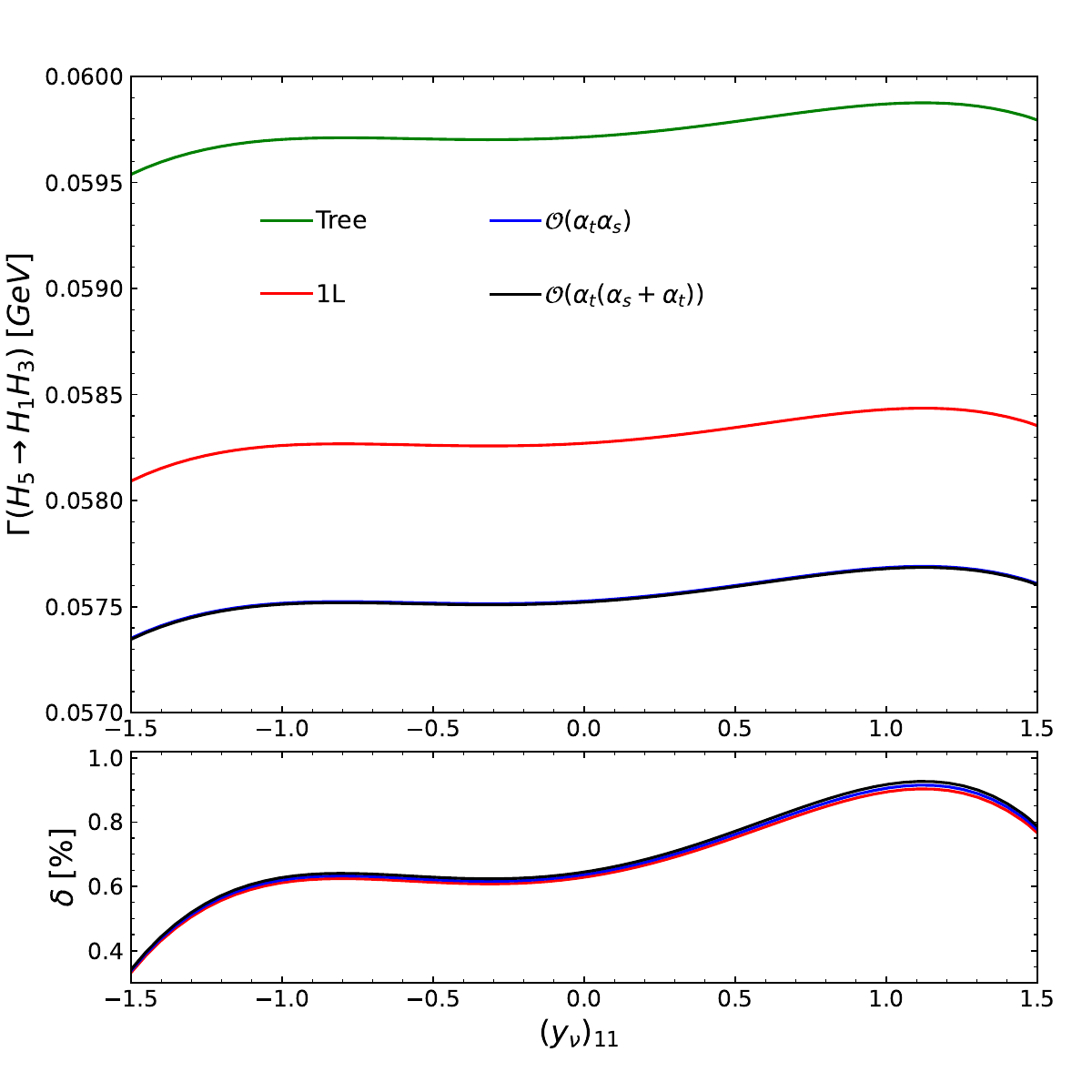} & 
\end{tabular}
\caption{ Upper panels: The decay widths at tree-level (green),
  one-loop (red),
  two-loop QCD ${\cal O}(\alpha_t \alpha_s)$ (blue) and
  two-loop EW ${\cal O}(\alpha_t (\alpha_t +
    \alpha_s))$ (black) for $H_2\to H_1H_1$, $H_4\to H_1H_1$, $H_4\to
  H_1H_2$, $H_4\to H_2H_2$, $H_5\to H_1H_3$  as functions of  the
  first component of the neutrino Yukawa matrix,
    $(y_\nu)_{11}$. Lower panels: Absolute values of the relative (s)neutrino corrections as defined in \eqref{eq:eqHOcorrections}.}
\label{fig:Yvwidths}
\end{figure}

We also investigate the dependence of the Higgs-to-Higgs decay widths on
parameters of the (s)neutrino sector  and
demonstrate the significant
effect coming from the neutrino Yukawa couplings. All other parameters
are found to give small
or mild effects. To illustrate our findings we present
in \figref{fig:Yvwidths} the dependence of the five decay widths
$\Gamma(H_2\to H_1H_1)$, $\Gamma(H_4\to H_1H_1)$, $\Gamma(H_4\to
H_1H_2)$, $\Gamma(H_4\to H_2H_2)$, and $\Gamma(H_5\to H_1H_3)$ on the
first component of the neutrino Yukawa matrix, $(y_\nu)_{11}$, which
is varied in the range $[-1.5,1.5]$. The  green, red, blue, and black
lines in the upper panels of each plot represent the decay widths at tree-level,
one-loop order, $\oorder{\alpha_t \alpha_s}$, and $\oorder{\alpha_t
(\alpha_s+\alpha_t)}$, respectively. We see that the tree-level decay
widths also depend on $(y_\nu)_{11}$.
This dependence stems from the WFR factors and the phase-space
factor and is quite noticeable for most of the 
considered decay channels. In
the lower panels of each plot, we show the absolute value of the
relative (s)neutrino  corrections defined as
 \be
 \delta^{x} =
 \abs{\frac{\Gamma^{x,\text{ISS}}-\Gamma^{x,\text{noISS}}}{\Gamma^{x-1,\text{ISS}}}} \label{eq:eqHOcorrections}
\ee
 with $x$ denoting the level of accuracy (1L/$\oorder{\alpha_t \alpha_s}$/$\oorder{\alpha_t
   (\alpha_s+\alpha_t)}$ corresponding to the (red/blue/black colors)
   and $x-1$ being the next-lower level of accuracy.
 With this definition, the nominator of \eqref{eq:eqHOcorrections} represents
 the (s)neutrino contributions to the width at  the considered order.
For all decays considered here, $\delta^{\text{1L}}$ is the largest
for the 
$H_2\to H_1 H_1$ decay channel, which involves
two SM-like Higgs bosons in the final states.
The correction to this particular channel is in the range $[7, 14.5]\%$ 
for the varied range of $(y_\nu)_{11}$, while it
remains below 5\% for the other decays. The qualitative behaviour of
$\delta$ 
 at $\oorder{\alpha_t \alpha_s}$ and 
$\oorder{\alpha_t(\alpha_s+\alpha_t)}$ is
quite similar to the one-loop one with different
magnitudes. This is because
 the nominators of \eqref{eq:eqHOcorrections} are quite close to the
 one-loop results, as the leading higher-order effects are
     independent of the ISS mechanism and therefore drop out, 
 while the higher-orders entering the denominator are different and
 not cancelled. For
 all plots in \figref{fig:Yvwidths}, the dependence of the widths and $\delta$ on  $(y_\nu)_{11}$ for the decay $H_5\to H_1 H_3$
  looks different than the ones for the other decays. In particular, the width of $H_5\to H_1 H_3$ is smallest at $(y_\nu)_{11}=-1.5$
  while it is the largest here for the other
  decays. This is due to the phase space factor $R_2$ which is
  smallest at $(y_\nu)_{11}=-1.5$. 
  
%%%%%%%%%%%%%%%%%%%%%%%%%%%%%%%%%%%%%%%%%%%%%%%%%%%%%%%%%%%%%
\section{Conclusions \label{sec:concl}}
%%%%%%%%%%%%%%%%%%%%%%%%%%%%%%%%%%%%%%%%%%%%%%%%%%%%%%%%%%%%%
In this paper we investigated the influence of the inverse seesaw
mechanism in the NMSSM for both the higher-order
corrections to the 
effective trilinear Higgs self-couplings and the Higgs-to-Higgs decays in comparison 
with the corrections on the Higgs boson masses. 
We computed the complete one-loop corrections to the Higgs-to-Higgs
decay widths taking into account non-vanishing external
momentum. Furthermore, we obtain the dominant
one-loop corrections from the (s)top sector and the (s)neutrino sector to the effective 
SM-like trilinear Higgs self-coupling computed in the limit of vanishing
external momenta and in the limit of vanishing gauge couplings. Regarding the renormalization, we
applied a mixed OS-$\overline{\mbox{DR}}$ scheme, which is the same 
scheme used for the loop-corrections to the Higgs boson masses. We then consistently
combined the one-loop result with the leading two-loop ${\cal
    O}(\alpha_t \alpha_s)$ and ${\cal O}(\alpha_t
    (\alpha_s + \alpha_t))$ results previously computed in the
    vanilla NMSSM by our group. In
the numerical analysis, we performed a parameter scan of the model and
kept only those points for our study that respect the constraints from the Higgs
data, the neutrino oscillation data, the charged lepton
flavor-violating decays $l_i \to l_j + \gamma$, and the new physics
constraints from the oblique parameters $S,T,U$.
We quantify the impact of the one-loop corrections from the (s)neutrino
sector on the SM-like Higgs trilinear
self-coupling and on the SM-like Higgs boson mass prediction by defining the 
relative differences $\Delta$ between the results with and without
(s)neutrino contributions. We find that the $\Delta$ for the
SM-like trilinear Higgs self-coupling can reach
10.5\% while the relative corrections to the SM-like Higgs boson
mass can reach reach 4.5\% for all found parameter
points. The relative effect of (s)neutrinos on the decay 
widths of the heavy Higgs decays can be significant for the decays of
the singlet-like and down-type-like
scalars decaying into a pair of the SM-like Higgs bosons.
For the Higgs-to-Higgs branching ratios 
of non-SM-like Higgs bosons, $\Delta$ is less than 3.4\% and mostly  encoded in
the external-leg corrections and phase-space factors.
%%%%%%%%%%%%%%%%%%%%%%%%%%%%%%%%%%%%%%%%%%%%%%%%%%%%%%%%%%%%%%
\section*{Acknowledgments}
M.M. acknowledges support by the Deutsche Forschungsgemeinschaft (DFG, German Research Foundation) under grant 396021762 - TRR 257.

\providecommand{\href}[2]{#2}\begingroup\raggedright\endgroup

\end{document}